\documentclass[10pt,journal,twoside]{IEEEtran}
\usepackage{multirow}
\usepackage{amssymb}
\usepackage[dvips]{graphicx}
\usepackage{amsmath}
\usepackage{amsthm}
\usepackage{latexsym,bm}
\usepackage{color}
\usepackage{subfigure}
\usepackage{longtable}
\usepackage{accents}
\usepackage{cite}
\usepackage{enumerate}
\usepackage{arydshln}
\usepackage{slashbox}
\usepackage{algorithmic}
\usepackage{booktabs}
\usepackage{subfigure}
\usepackage{stfloats}
\usepackage[table,xcdraw]{xcolor}
\usepackage{graphicx}
\usepackage{bm}

\makeatletter
\def\widebar{\accentset{{\cc@style\underline{\mskip10mu}}}}
\def\Widebar{\accentset{{\cc@style\underline{\mskip8mu}}}}
\makeatother

\theoremstyle{plain}

\theoremstyle{definition}
\theoremstyle{definition}

\begin{document}
\title{{Design and Performance Analysis of a New STBC-MIMO LoRa System}}

\author{\fontsize{11pt}{\baselineskip}\selectfont{Huan Ma, Guofa Cai, Yi Fang, Pingping Chen, Guojun Han}
\thanks{H. Ma, G. Cai, Y. Fang and G. Han are with the School of Information Engineering, Guangdong University of Technology, Guangzhou 510006, China (e-mail: mh-zs@163.com; caiguofa2006@126.com; fangyi@gdut.edu.cn; gjhan@gdut.edu.cn).}
\thanks{P.~Chen is with the Department of Electronic Information, Fuzhou University, Fuzhou 350116, China (e-mail: ppchen.xm@gmail.com).}}

\maketitle
\begin{abstract}
LoRa is a modulation technology for low power wide area networks (LPWAN) with enormous potential in 5G era. However, the performance of LoRa system deteriorates seriously in fading-channel environments. To tackle this problem, in this paper we introduce multiple-input-multiple-output (MIMO) configuration employing space-time block coding (STBC) schemes into the LoRa system to formulate an STBC-MIMO LoRa system. Then, we investigate the theoretical performance of the proposed system over Rayleigh fading channels. To this end, we derive the distribution of the decision metric for the demodulator in the proposed system. Based on the above distribution, we propose the closed-form approximated bit error rate (BER) expression of the proposed system when perfect and imperfect channel information states (CSIs) are considered. In addition, we analyze the diversity order of the proposed system. The result demonstrates that the diversity order of the system in the imperfect CSI scenario with fixed channel estimate error variance is zero. However, in the imperfect CSI scenario with a decreasing channel estimate error variance and the perfect CSI scenario, the system can achieve full diversity.
Finally, experimental results verify the accuracy of the theoretical analysis and the excellent performance of the proposed system. Due to such superiority, the proposed STBC-MIMO LoRa system can be considered as a good scheme for LPWAN.

\end{abstract}
\begin{IEEEkeywords}
Internet of things (IoT), LoRa, bit error rate (BER), space-time block coding (STBC), multiple-input-multiple-output (MIMO), diversity order, Rayleigh fading channel.
\end{IEEEkeywords}

\section{Introduction}
With the continuous update and improvement of Internet of things (IoT) and 5G technologies, massive connectivity has become one of the most important backbone of current world development. By providing real-time information, analysis and decision-making, the IoT in the 5G era is greatly changing the previous business models and lifestyles of people, such as driverless cars, unmanned aerial vehicle, smart appliances, and automated factories, thereby bringing unlimited possibilities for the future development of society. The widely promoted low power wide area network (LPWAN) plays an important role in 5G networks in recent years \cite{7397856}, because it can complement traditional cellular and short-range wireless technologies to address the different requirements of IoT applications. It is estimated that in 2022, the number of compatible nodes for LPWAN will reach 350 million \cite{9000820}.

For LPWAN, several communication technologies have been proposed, such as NB-IoT \cite{8120239}, Sigfox \cite{7925650}, and LoRa \cite{8268120,7815384}. Compared with NB-IoT- and Sigfox-based LPWAN, LoRa-based LPWAN has some unique advantages such as highly open and flexible. Furthermore, LoRa-based LPWAN can be deployed in a private network and the operating cost of this LPWAN is low.
As a modulation scheme of LPWAN, LoRa has gained considerable commercial traction and its specifications are maintained by the LoRa Alliance.\footnote{https://lora-alliance.org}
LoRa is a low-power, low-speed, and long-range modulation based on chirp spread-spectrum (CSS) technology \cite{8886735}.
In the LoRa modulation, cyclic shifts of chirp signal with linearly increased frequency form a multidimensional space for LoRa symbols, and the signals of different LoRa symbols are orthogonal to each other \cite{8067462}.
The coverage of LoRa is determined by the spreading factor. Increasing the spreading factor can provide wider coverage but reduce the data rate \cite{8633860}.

With the expansion of the market share occupied by LoRa in LPWAN, the number of countries deploying LoRa-based solutions has grown rapidly, which has reached 142 \cite{7815384}.
Accordingly, LoRa has also attracted more and more attention from academia.
To explore various performance indicators of the LoRa modulation, researchers have carried out a lot of experimental-based works in the real world. In \cite{8288154,7803607,7577098,7377400}, the coverage capability of LoRa has been studied.
In different scenarios, the coverage of LoRa ranges from 100m to 30km. In \cite{7577098}, the coverage of LoRa in outdoor scenario has been evaluated by deploying LoRa base station on a mountain.
The results of the study indicate that the LoRa base station at an altitude of 470m can cover an area of 1380 square kilometers. In \cite{7377400}, it has been observed that the available communication distances of LoRa are 15km in the ground environment and 30km in the water environment.

After some pioneering work based on experiments, research on LoRa networks has attracted growing interest. With the increasing density of IoT applications, some work has considered the scalability and capacity of LoRa. In \cite{7974300}, the mathematical tools has been utilized to simulate the uplink coverage of the single gateway LoRa network and it reveals the unique physical layer characteristics of the LoRa network. In \cite{8090518}, scalability performance has been conducted on a LoRa network with multiple gateways, and the impact of confirmed and unconfirmed messages on the traffic of a large-scale LoRa network has been analyzed.
In addition, theoretical LoRa capacity has been explored in \cite{12444458888,7499263,8903531}, where ALOHA network model is utilized to characterize LoRa-based LPWAN. In \cite{8839056}, a more practical model considering capture effect has been adopted to analyze LoRa capacity.
Moreover, some recent research works have been studied to design multi-hop and multi-relay schemes for improving reliability of LoRa networks, e.g., concurrent transmission multi-hop LoRa network \cite{8048465}, multi-hop LoRa network with tree-based spreading factor clustering algorithm \cite{8637935}, and cooperative LoRa networks \cite{9016483,9014071}.

While the LoRa physical layer is protected by patent \cite{Seller2016Low}, the study on such a modulation has become more active recently due to the implementation of reverse engineering \cite{grcon,77566824}. A rigorous mathematical description of the LoRa modulation and demodulation process has been developed in \cite{8067462}, and subsequently the waveform and spectral characteristics of the LoRa modulation have been investigated in \cite{8723130}. In addition, bit-error-rate (BER) performance analysis of the LoRa modulation has been performed in \cite{8392707,sensosloraber}. The analytical results show that the poor performance of the LoRa system might not sustain long-range communication in the fading environment.
For this reason, some improved modulation schemes based on the physical layer of LoRa have been proposed, such as phase-shift-keying (PSK) LoRa \cite{8746470}, interleaved chirp spreading LoRa \cite{8607020}, and dual orthogonal LoRa \cite{8937880}. All of these schemes aim at enhancing the capacity of the LoRa network.
A LoRa-aided binary PSK multiple-input multiple-output (MIMO) system has been presented in \cite{9018816}. The MIMO LoRa system can achieve better BER performance compared with the single-input single-output (SISO) LoRa system. However, the LoRa symbol in this system is only utilized to spread the spectrum of the BPSK baseband signal rather than carrying information, which leads to severe data-rate loss.

With the above motivation and considerable scalability of LoRa\cite{7815384}, this paper propose a space-time block-coded (STBC) MIMO scheme\cite{9018816,8746692,8396085,8007277} with the LoRa modulation. The contributions of this paper are summarized as follows:
\begin{enumerate}[1)]
\item \label{}
 An STBC-MIMO LoRa system with $M$ transmit antennas and $N$ receive antennas is put forward. The proposed system can greatly improve BER performance, thereby enhancing the reliability of LoRa networks in fading-channel environments.
\item \label{}
The theoretical BER performance of the proposed system is carefully analyzed over Rayleigh fading channels. Based on the principle of the STBC-MIMO scheme, the theoretical model of the proposed system is established, then the distribution of the decision metric for the demodulator of the proposed system is derived. According to the above distribution, the closed-form approximated BER expression of the proposed system is presented for both perfect and imperfect channel information states (CSIs). In particular, two common channel estimation error models (CEEMs) are considered in the analysis.
\item \label{}
The asymptotic BER performance is investigated to analyze the diversity order of the proposed system. The results indicate indicates that for the fixed channel estimation error variance, the system reaches zero diversity order, while for the perfect CSI and channel estimation error variance being a decreasing function of average signal-to-noise ratio (SNR), the system can achieve full diversity $d=MN$.
\item \label{}
Simulation results not only demonstrate the superior performance of the proposed STBC-MIMO LoRa system, but also verify the accuracy of the approximated BER expressions and diversity-order analysis.
\end{enumerate}

The remainder of the paper is organized as follows. In Section~\ref{sect:system model}, we provide the detailed descriptions of the LoRa modulation/demodulation process and the proposed STBC-MIMO LoRa system. In Section~\ref{sect:performance analysis}, we present the closed-form approximated BER expressions of the proposed system for perfect and imperfect CSIs. We also analyze the diversity order of the proposed system in the same Section. In Section~\ref{sect:simulation results and discussion}, we present various simulation results with some discussions. Finally, Section~\ref{sect:Conclusions} concludes the paper.

\section{System Model}\label{sect:system model}
\subsection{LoRa Modulation}\label{sect:lora modulation}
LoRa is a frequency shift CSS based modulation scheme. In the LoRa modulation, the frequency of the baseband signal varies linearly in a symbol duration and the bandwidth of the LoRa signal is ${B_w}$.
There are ${2^{{{SF}}}}$ chips in each LoRa symbol, where ${{SF}} \in \left\{ {7,8,...,12} \right\}$ is the spreading factor of LoRa.
For a LoRa symbol ${x_o }$, it can carry ${{SF}}$ bits, and assuming that the ${o ^{th}}$ transmitted symbol is ${s_o } = p \in \left\{ {0,1,...,{2^{{{SF}} - 1}}} \right\}$. The frequency of ${x_o }$ varies linearly from the starting frequency ${f_s} = \frac{{{B_w} \cdot p}}{{{2^{SF}}}}$ to ${B_w}$ and then folds to $0$, in the remaining symbol duration, the frequency continues to change linearly from $0$ to ${f_s}$ \cite{8903531}. Specifically, the frequency of each chip increases by $\frac{{{B_w}}}{{{2^{SF}}}}$.
Therefore, the discrete-time baseband signal of the LoRa symbol ${x_o }$ can be expressed as \cite{8903531}
\begin{align}
  &{w_o }\left( {\kappa{T_c}} \right) = \sqrt {{E_s}} {{\bar w}_p}\left( {\kappa{T_c}} \right) \hfill \nonumber\\
  &= \sqrt {\frac{{{E_s}}}{{{2^{SF}}}}} \exp \left[ {j2\pi \left( {\frac{{{{\left( {\left( {p + \kappa} \right)\bmod {2^{SF}}} \right)}^2}}}{{{2^{SF{\text{ + 1}}}}}}} \right)} \right], \hfill
  \label{eq:1func}
\end{align}
where ${T_c} = \frac{1}{{{B_w}}}$ is the sample interval, $\kappa$ denotes the index of the sample at time $\kappa{T_c}$, ${{E}_{s}}$ is the symbol energy, and ${\bar w_p}(\kappa{T_c})$ is the basis function of ${w_o }(\kappa{T_c})$. As seen from Eq.~(\ref{eq:1func}), the LoRa signal transmitting the symbol $p$ can be considered as a cyclic shift of $p{{T}_{c}}$ for the basis CSS signal \cite{8903531,8392707}.\footnote{The basic CSS signal can also be called as the \textit{upchirp} signal, and its frequency varies linearly from $0$ to ${{B}_{w}}$ in a symbol duration.} Since chirp signals with different offsets are mutually orthogonal, for the LoRa signal of the symbol ${s_o }$, when it correlates with ${{2}^{{SF}}}$ possible LoRa signals, it has the following properties \cite{8392707}
\begin{align}
{\Lambda _i} = \sum\limits_{\kappa = 0}^{{2^{{SF}}} - 1} {{w_o }\left( {\kappa{T_c}} \right)}  \cdot \bar w_i^*\left( {\kappa{T_c}} \right) = \left\{ \begin{gathered}
  \sqrt {{E_s}}, {\text{    }}i = p \hfill \\
  0,{\text{         }}\ \ \ \,\,\,\,   i \ne p \hfill
\end{gathered}  \right.,
\label{eq:2func}
\end{align}
where $0\le i\le {{2}^{{SF}}}-1$ and $ * $ is the complex conjugate operation. The demodulation of the LoRa signal can be performed based on the above properties. For a received signal ${r_o }\left( {\kappa{T_c}} \right)$ of the LoRs symbol ${x_o }$ after transmission over a frequency-flat and time-invariant channel, the output of the correlator in the LoRa demodulator is written as\cite{8392707}
\begin{align}
  &{{\dot \Lambda }_i} = \sum\limits_{\kappa = 0}^{{2^{SF}-1}} {{r_o }\left( {\kappa{T_c}} \right) \cdot \bar w_i^*\left( {\kappa{T_c}} \right)}  \hfill \nonumber\\
  & = \sum\limits_{\kappa = 0}^{{2^{SF}-1}} {\left( {\sqrt {{h_c}} {w_o }\left( {\kappa{T_c}} \right) + {w_n}(\kappa{T_c})} \right)}  \cdot \bar w_i^*\left( {\kappa{T_c}} \right) \hfill \nonumber\\
  &= \left\{ \begin{aligned}
  &\sqrt {{h_c}{E_s}}  + {w_{n,i}},{\text{      }}i = p \hfill \\
  &{w_{n,i}},\ \ \ \ \ \ \ \ \ \ \ \ \ \, i \ne p \hfill
\end{aligned}  \right.,
\label{eq:3func}
\end{align}
where $\sqrt{{{h}_{c}}}$ is the complex envelope amplitude, ${w_n}\left( {\kappa{T_c}} \right)$ is the complex additive Gaussian white noise (AWGN), and ${{w}_{n,i}}$ is the corresponding complex Gaussian noise process \cite{8607020}. Hence, the symbol ${s_o }$ can be estimated as
\begin{align}
{\hat s_o } = \arg \mathop {\max }\limits_{i = 0,...,{2^{SF}-1}} \left( {\left| {{{\dot \Lambda }_i}} \right|} \right),
\label{eq:4func}
\end{align}
where $\left| \cdot \right|$ denotes absolute operation.
In addition, another equivalent low complexity method can also be utilized for demodulation. First, the received signal is multiplied with downchirp ${\bar w_{down}}\left( {\kappa{T_c}} \right)$ (this step is called a \textit{dechirping}) \cite{8903531}, where ${\bar w_{down}}\left( {\kappa{T_c}} \right)$ can be expressed as
\begin{align}
{\bar w_{down}}\left( {\kappa{T_c}} \right) = \sqrt {\frac{1}{{{2^{SF}}}}} \exp \left( { - j2\pi \frac{{{\kappa^{2}}}}{{{2^{SF + 1}}}}} \right).
\label{eq:5func}
\end{align}
Afterwards, the ${{2}^{{SF}}}-\text{point}$ discrete Fourier transform (DFT) is performed on the dechirped signal, thus one can obtain as
\begin{align}
{{\dot{\bm{\Lambda}}}_{F}} = {\text{DFT}}\left( {{\textbf{r}_o } \odot {{\bar{\textbf{w}}}_{down}}} \right),
\label{eq:6func}
\end{align}
where ${{\dot{\bm{\Lambda }}}_{F}}\!=\!\left[ {{{\dot{\Lambda }}}_{F,0}},...{{{\dot{\Lambda }}}_{F,i}},...{{{\dot{\Lambda }}}_{F,{{2}^{{SF}}}-1}} \right]$, ${{\textbf{r}}_{{{o }}}}=\left[ {{r}_{{{o }}}}\left( 0 \right),{{r}_{{{o }}}}\left( {{T}_{c}} \right),...,{{r}_{{{o }}}}\left( ({{2}^{{SF}}}-1){{T}_{c}} \right) \right]$, ${{\bar{\textbf{w}}}_{down}}=\left[ {{{\bar{w}}}_{down}}\left( 0 \right),{{{\bar{w}}}_{down}}\left( {{T}_{c}} \right),...,{{{\bar{w}}}_{down}}\left( \left( {{2}^{{SF}}}-1 \right){{T}_{c}} \right) \right]$, and $\odot $ is the Hadamard product operator \cite{Magnus1988Matrix}. Next, the LoRa symbol is estimated by selecting the index of the frequency bin with the maximum magnitude, given by
\begin{align}
{\hat s_o } = \arg \mathop {\max }\limits_{i = 0,...,{2^{SF}} - 1} \left( {\left| {{{\dot \Lambda }_{F,i}}} \right|} \right).
\label{eq:7func}
\end{align}
\begin{figure}[tbp]
\center
\subfigure[\hspace{-0.0cm}]{ \label{fig:subfig:1a}
\includegraphics[width=3.3in,height=2.848596in]{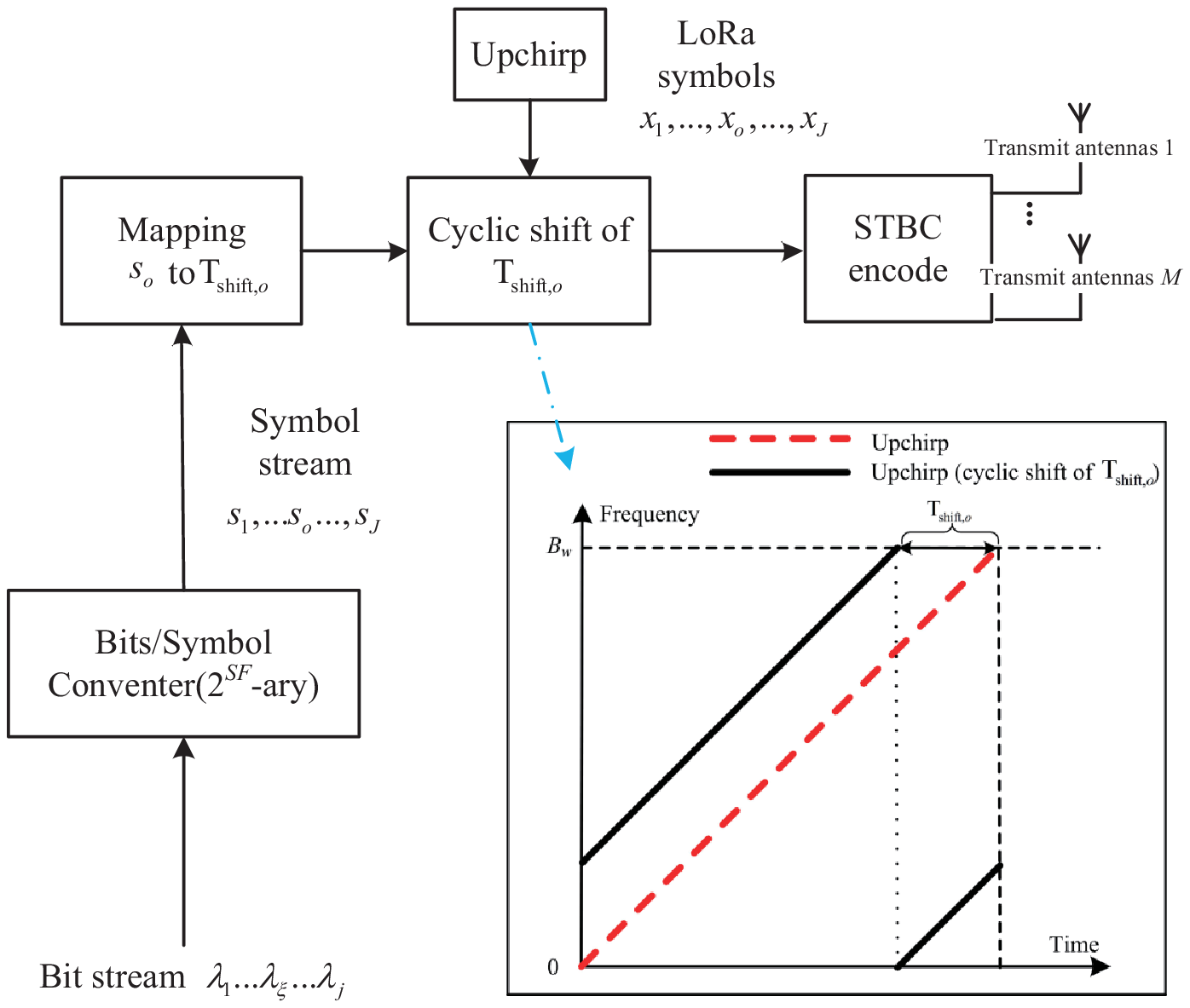}}
\subfigure[\hspace{-0.0cm}]{ \label{fig:subfig:1b}
\includegraphics[width=3.3in,height=1.687357in]{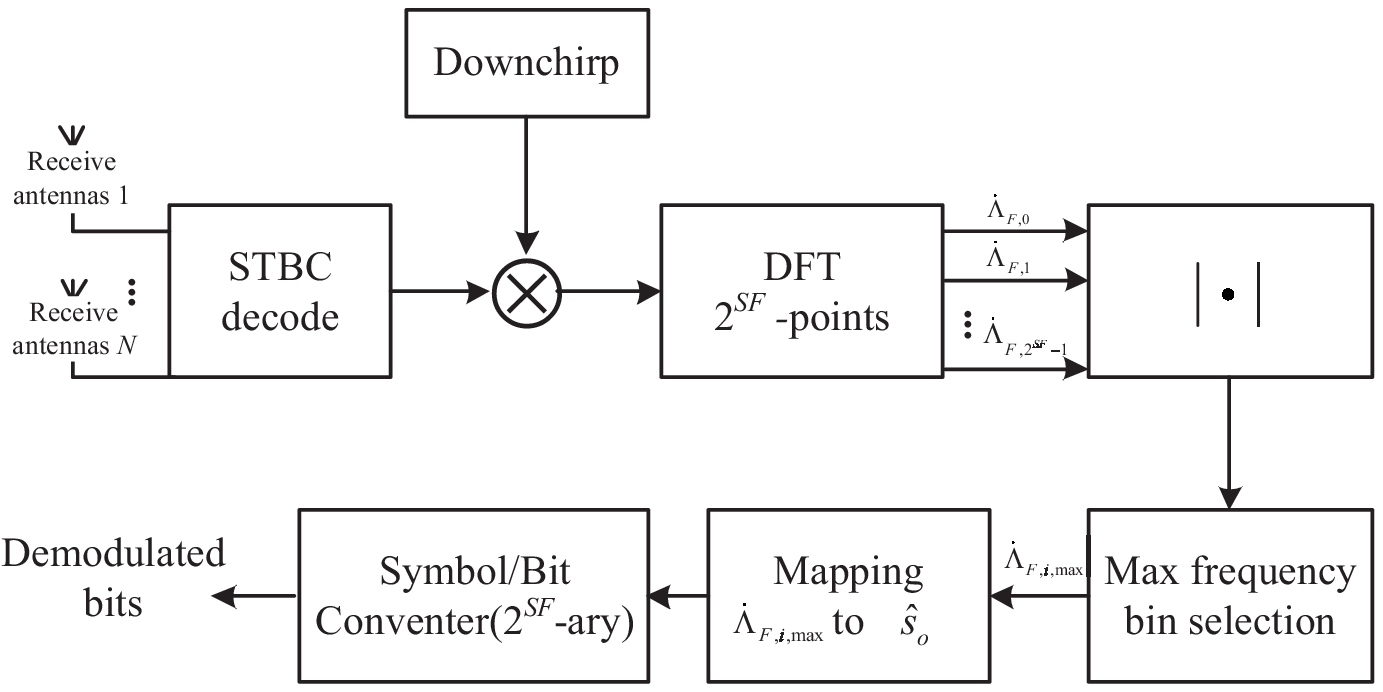}}
\caption{Illustration of one possible realization of (a) transmitter and (b) receiver for the proposed STBC-MIMO LoRa system.}
\label{fig:fig1}
\vspace{-2mm}
\end{figure}
\subsection{The Proposed STBC-MIMO LoRa System}\label{sect:stbc-mimo lora system}
In this paper, we consider a wireless MIMO system with $M$ transmit antennas and $N$ receive antennas that operates over a flat and quasi-static Rayleigh fading channel, therefore the path gains remain constant in a frame of $J$ symbols and varies from a frame to another. We represent the MIMO channel as an $M\times N$ matrix, $\textbf{H} = \left\{ {{h_{m,n}}} \right\}$, where ${h_{m,n}}$ denotes the complex channel gain from the ${m^{th}}$ transmit antenna to the ${n^{th}}$ receive antenna. ${h_{m,n}}$ are independent random variables that follow complex Gaussian distribution with zero mean and variance $0.5$ per dimension.
Fig.~\ref{fig:fig1} illustrates one possible realization on the transmitter and receiver of the proposed STBC-MIMO LoRa system, it is worth noting that the channel estimation module is in the STBC decoder structure.\footnote{Since the LoRa system can be regarded as a chirp-modulated high-order ${2^{{{SF}}}}$--ary FSK system, the proposed STBC-MIMO LoRa system can adopt the pilot-based channel estimation method which is applicable to the FSK system \cite{6192277,8746311}.} A complex orthogonal STBC transmission matrix $\textbf{G}$ is represented by a $U \times M$ transmission matrix, where the entries are linear combinations of ${g_1},{g_2},...,{g_J}$ and their conjugates. Moreover, matrix $\textbf{G}$ satisfies complex orthogonality ${\textbf{G}^H}\textbf{G} = {u_{cons}}\left( {{{\left| {{g_1}} \right|}^2} + ... + {{\left| {{g_J}} \right|}^2}} \right){\textbf{I}_M}$\cite{7711466,1275699}, where ${\textbf{I}_M}$ is an $M \times M$ identity matrix and ${u_{cons}}$ is a constant that depends on the STBC transmission matrix \cite{6400454}. Matrix $\textbf{G}$ is utilized to encode $J$ $\left( {J  \geqslant 2} \right)$ input symbols into an $M$-dimensional vector sequence of $U$ time slot (i.e., to control the symbol transmission of $M$ transmitting antennas in each slot). Consequently, the transmission rate of STBC is $r = J/U$.

In this paper, the channel estimation matrix at the receiver is modeled as \cite{8316922,8016612,8985541}
\begin{align}
\hat {\textbf{H}} = \textbf{H} + {\textbf{E}_h},
\label{eq:8func}
\end{align}
where ${\textbf{E}_h} = \left\{ {{e_{m,n}}} \right\}$ is an $M \times N$ error matrix independent of $\textbf{H}$, ${e_{m,n}}$ are complex Gaussian independent random variables with zero mean and variance $\sigma _e^2$, where $\sigma _e^2$ reflects the accuracy of the channel estimation. Accordingly, the variance of the estimated channel gain is $\sigma _{\hat h}^2 = \sigma _h^2 + \sigma _e^2$. In particular, when $\sigma _e^2 = 0$, it can be considered that perfect channel estimation is performed at the receiver (i.e., the receiver knows perfect CSI). In this paper, two types of channel estimation error variance are considered: one is that $\sigma _e^2$ is fixed and independent of the average SNR, and the other is that $\sigma _e^2$ is a decreasing function of the average SNR. These two types of error variance correspond to CEEMs I and II, respectively. For the proposed STBC-MIMO LoRa system, when the channel estimation error occurs at the receiver, it causes inter-antenna interference (IAI). Here, we take the STBC ${\textbf{G}_2}$ in \cite{753730} as an example and apply it to an STBC-MIMO LoRa system with two receive antennas to illustrate the encoding/decoding process in the proposed system. Code ${\textbf{G}_2}$ is given by
\begin{align}
{\textbf{G}_2} = \left( {\begin{array}{*{20}{c}}
  {{g_1}}&{{g_2}} \\
  { - g_2^*}&{g_1^*}
\end{array}} \right).
\label{eq:9func}
\end{align}

Utilizing ${\textbf{G}_2}$, we can encode the LoRa symbol in space and time. Taking the encoding of the first two LoRa symbols ${x_1}$ (the symbol transmitted by ${x_1}$ is ${s_1} = p$) and ${x_2}$ (the symbol transmitted by ${x_2}$ is ${s_2} \ne p$) in the sequence as an example, this process is shown in Table~\ref{tablei}.
Correspondingly, the notation for the received signal at the two receive antennas is defined in Table~\ref{tableii},
\begin{table}[t]
\caption{The Encoding and Transmission Sequence for the proposed STBC-MIMO LoRa system with two Transit Antennas}
\centering\vspace{0mm}
\begin{tabular}{|c|c|c|}
\hline
   & {Transmit antenna 1} & {Transmit antenna 2} \\ \hline
{Time ${t_s}$} & {${x_{1}}$}                 & {${x_{2}}$}                 \\ \hline
{Time ${t_s} + {2^{SF}}{T_c}$} & {$ - x_{2}^ * $}                 & {$x_{1}^ * $}                 \\ \hline
\end{tabular}
\label{tablei}
\vspace{0mm}
\end{table}
\begin{table}[t]
\caption{The Notation for the Received Signal at the Two Receive Antennas}
\centering\vspace{0mm}
\begin{tabular}{|c|c|c|}
\hline
   & {Receive antenna 1} & {Receive antenna 2} \\ \hline
{Time ${t_s}$} & {${r_1}$}                 & {${r_3}$}                 \\ \hline
{Time ${t_s} + {2^{SF}}{T_c}$} & {${r_2}$}                 & {${r_4}$}                 \\ \hline
\end{tabular}
\label{tableii}
\vspace{0mm}
\end{table}
where
\begin{align}
 & {r_1} = {h_{1,1}}{x_1} + {h_{2,1}}{x_2} + {n_1} \hfill \nonumber\\
  &{r_2} =  - {h_{1,1}}x_2^ *  + {h_{2,1}}x_1^ *  + {n_2} \hfill \nonumber\\
  &{r_3} = {h_{1,2}}{x_1} + {h_{2,2}}{x_2} + {n_3} \hfill \nonumber\\
  &{r_4} =  - {h_{1,2}}x_2^ *  + {h_{2,2}}x_1^ *  + {n_4}, \hfill
  \label{eq:9funcadd1}
\end{align}
and ${n_1}$, ${n_2}$, ${n_3}$, and ${n_4}$ are the complex AWGN with variance ${N_0}/2$ per dimension.
Next, the maximum likelihood decoding of the STBC can be achieved by linear processing \cite{753730} in the STBC decoder at the receiver, thereby recovering the desired LoRa symbols. Without loss of generality, we take ${x_1}$ as an example for illustration and following analysis. For the LoRa symbol ${x_1}$, LoRa symbols that carrying other transmitted symbols (i.e., the symbol is not equal to $p$) can be regarded as interference. After STBC decoding, the recovered LoRa symbol ${\tilde x_1}$ can be expressed as
\begin{align}
  {\tilde x_1} &= \hat h_{1,1}^ * {r_1} + {{\hat h}_{2,1}}r_2^ *  + \hat h_{1,2}^ * {r_3} + {{\hat h}_{2,2}}r_4^ *  \hfill \nonumber\\
   &= \left( {{{\left| {{{\hat h}_{1,1}}} \right|}^2} + {{\left| {{{\hat h}_{2,1}}} \right|}^2} + {{\left| {{{\hat h}_{1,2}}} \right|}^2} + {{\left| {{{\hat h}_{2,2}}} \right|}^2}} \right. \hfill \nonumber\\
  &\underbrace {\left. { - \hat h_{1,1}^*{e_{1,1}} - \hat h_{2,1}^*{e_{2,1}} - {{\hat h}_{1,2}}e_{1,2}^* - {{\hat h}_{2,2}}e_{2,2}^*} \right){x_1}}_{{S_\alpha }} \hfill\nonumber \\
   &+ \underbrace {\left( { - \hat h_{1,1}^ * {e_{1,2}} - \hat h_{2,1}^ * {e_{2,2}} + {{\hat h}_{1,2}}e_{1,1}^ *  + {{\hat h}_{2,2}}e_{2,1}^ * } \right){x_2}}_{{S_\beta }} \hfill \nonumber\\
   &+ \underbrace {\left( {\hat h_{{\text{1,1}}}^ * {n_1} + \hat h_{2,1}^ * {n_2} + {{\hat h}_{1,2}}n_3^ *  + {{\hat h}_{2,2}}n_4^ * } \right)}_{{S_\tau }}, \hfill
  \label{eq:10func}
\end{align}
where terms ${{S}_{\alpha }}$, ${{S}_{\beta }}$, and ${{S}_{\tau }}$ are the desired signal, IAI and noise, respectively. Accordingly, referring to Eq.~(\ref{eq:3func}) and Eq.~(\ref{eq:4func}), the decision metric of ${\tilde x_1}$ can be expressed as
\begin{align}
&{Z_{{{\tilde x}_{1}},i}} = \left| {\sum\limits_{{\kappa} = 0}^{{2^{{{SF}}}-1}} {{w_{{{\tilde x}_1}}}\left( {{\kappa}{T_c}} \right) \cdot \bar w_{i}^*\left( {{\kappa}{T_c}} \right)} } \right|\nonumber\\
& =\!\! \left\{ \begin{array}{l}
\!\!\!\left| \begin{array}{l}
\!\!\sqrt {\frac{{{E_s}}}{2}} \left( {{{\left| {{{\hat h}_{1,1}}} \right|}^2} + {{\left| {{{\hat h}_{2,1}}} \right|}^2} + {{\left| {{{\hat h}_{1,2}}} \right|}^2} + {{\left| {{{\hat h}_{2,2}}} \right|}^2}} \right.\\
\!\!\left. { - \hat h_{1,1}^ * {e_{1,1}} - \hat h_{2,1}^ * {e_{2,1}} - {{\hat h}_{1,2}}e_{1,2}^ *  - {{\hat h}_{2,2}}e_{2,2}^ * } \right)\\
\!\! + \hat h_{1,1}^ * {\phi _1} + \hat h_{2,1}^ * {\phi _2} + {{\hat h}_{1,2}}\phi _3^ *  + {{\hat h}_{2,2}}\phi _4^ *
\end{array} \right|,{\rm{     }}\ \ \ i = p\\
\!\!\!\left| \begin{array}{l}
\sqrt {\frac{{{E_s}}}{2}} \left( { - \hat h_{1,1}^ * {e_{1,2}} + {{\hat h}_{1,2}}e_{1,1}^ *  - \hat h_{2,1}^ * {e_{2,2}}} \right.\\
\left. { + {{\hat h}_{2,2}}e_{2,1}^*} \right) + \hat h_{1,1}^ * {\phi _1} + \hat h_{2,1}^ * {\phi _2}\\
 + {{\hat h}_{1,2}}\phi _3^ *  + {{\hat h}_{2,2}}\phi _4^ *
\end{array} \right|,{\rm{    }}i \ne p,i = {s_2}\\
\!\!\!\left| {\hat h_{1,1}^ * {\phi _1} + \hat h_{2,1}^ * {\phi _2} + {{\hat h}_{1,2}}\phi _3^ *  + {{\hat h}_{2,2}}\phi _4^ * } \right|,{\rm{        }}\ \ \ \ \ \, i \ne p,i \ne {s_2}
\end{array} \right.,
\label{eq:11func}
\end{align}
where ${{\phi }}$ is the complex Gaussian noise process. Then, the symbol ${{s}_{1}}$ is estimated by
\begin{equation}
{{\hat{s}}_{1}}=\arg \underset{i=0,...,{{2}^{{SF}}-1}}{\mathop{\max }}\,\left( \left| {Z_{{{\tilde x}_{1}},i}} \right| \right).
\label{eq:12func}
\end{equation}
\section{Performance Analysis}\label{sect:performance analysis}
In this section, the average BER performance of the proposed STBC-MIMO LoRa system is analyzed. We denote ${{f}_{Ray}}\left( y;{{\sigma }_{y}} \right)$ and ${{f}_{Ri}}\left( y;{{m}_{y}},{{\sigma }_{y}} \right)$ as the probability density functions (PDFs) of the Rayleigh and Rice distributions, respectively, and we denote the cumulative probability density (CDF) of Rayleigh distribution by ${{F}_{Ray}}\left( y;{{\sigma }_{y}} \right)$, where ${{m}_{y}}$ and ${{\sigma }_{y}}$ are the scale and location parameters of the variable $y$ \cite{John1983Digital}. By induction on Eq.~(\ref{eq:11func}), we obtain the distribution of the decision metric of ${x_{1}}$ for an STBC-MIMO LoRa system with $M$ transmit antennas and $N$ receive antennas
\begin{align}
{Z_{{{\tilde x}_{1}},i}}\!\! \sim\!\! \left\{ \begin{array}{l}
\!\!\!\!{f_{Ri}}\!\!\left(\!\!{\alpha ;\!||\hat{\textbf{H}}||_F^2\!\!\sqrt {\frac{{{E_s}}}{{rM}}} ,\sqrt {\frac{||\hat{\textbf{H}}||_F^2}{2}\left( {\frac{{\sigma _e^2{E_s}}}{{rM}} + {N_0}} \right)} } \right),{\rm{     }}i = p\\
\!\!\!\!{f_{Ray}}\!\!\left(\!\!{\beta ;\!\!\sqrt {\frac{{||\hat{\textbf{H}}||_F^2}}{2}\!\!\left( {\frac{{\sigma _e^2{E_s}}}{{rM}}\!\!+\!\!{N_0}} \right)} } \right),{\rm{                      }}i \ne p,i = {s_2},...,{s_J}\\
\!\!\!\!{f_{Ray}}\!\!\left( {\tau ;\sqrt {\frac{{||\hat{\textbf{H}}||_F^2{N_0}}}{2}} } \right),{\rm{                                       }}\ \ \ \ \ \ \ \ \ \,i \ne p,{s_2},...,{s_J}
\end{array} \right.,
\label{eq:13func}
\end{align}
where $||\hat{\textbf{H}}||_F^2 = \sum\nolimits_{m = 1}^M {\sum\nolimits_{n = 1}^N {{{\left| {{\hat{h}_{m,n}}} \right|}^2}} }$ is the square of the Frobenius norm of $\left\{ {{\hat{h}_{m,n}}} \right\}$. For convenience, we denote $||\hat{\textbf{H}}||_F^2$ by $X$. For a Rayleigh fading channel, $X$ follows a chi-square distribution with $MN$ degrees of freedom. Thus, the average BER of the proposed system can be expressed as
\begin{figure*}[tbp]
\centering
\subfigure[\hspace{-0.0cm}]{\label{fig:subfig:2a}
\begin{minipage}[t]{2.25in}
\centering
\includegraphics[width=2.25in]{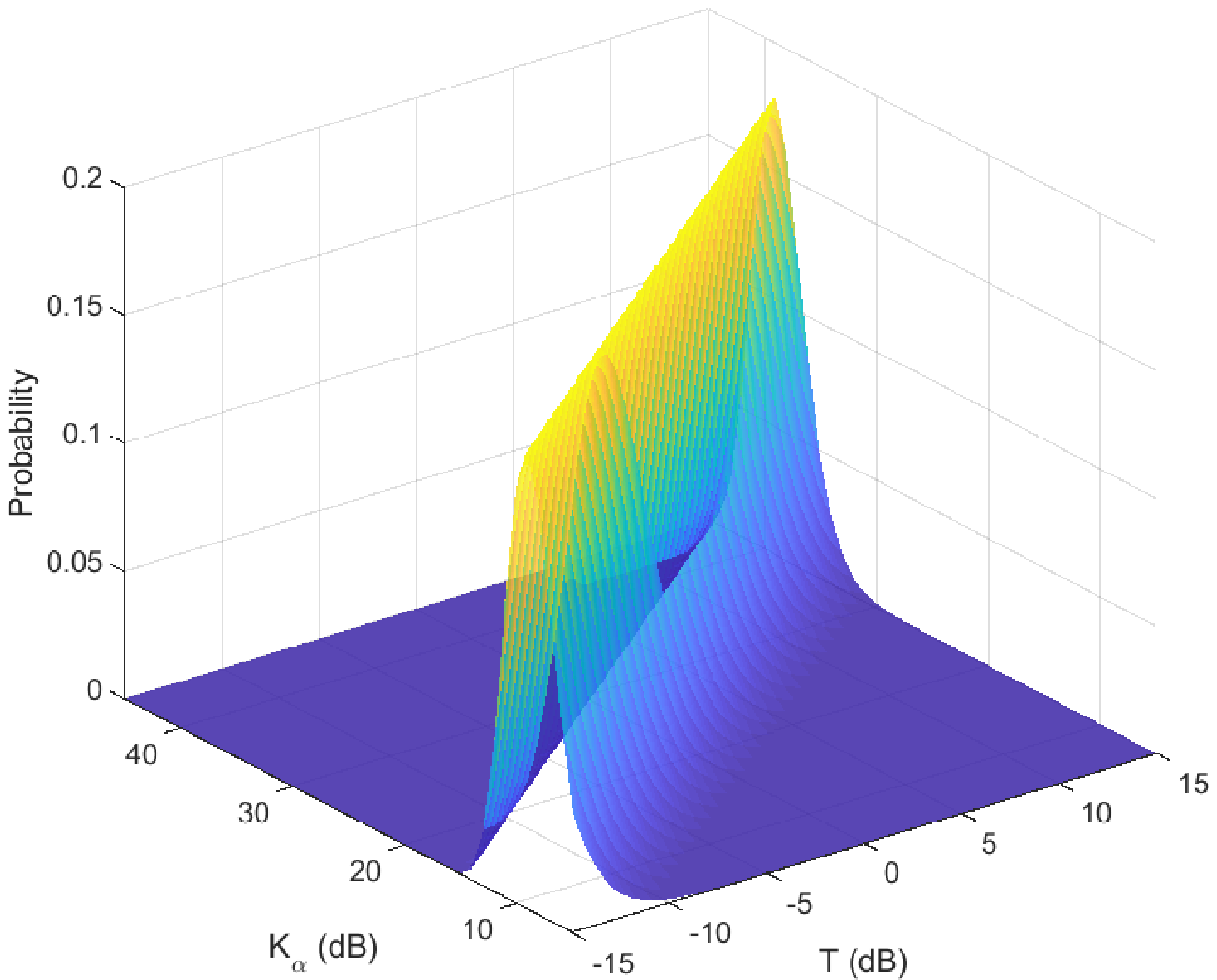}
\vspace{0.00cm}
\end{minipage}%
}%
\subfigure[\hspace{-0.0cm}]{\label{fig:subfig:2b}
\begin{minipage}[t]{2.25in}
\centering
\includegraphics[width=2.25in]{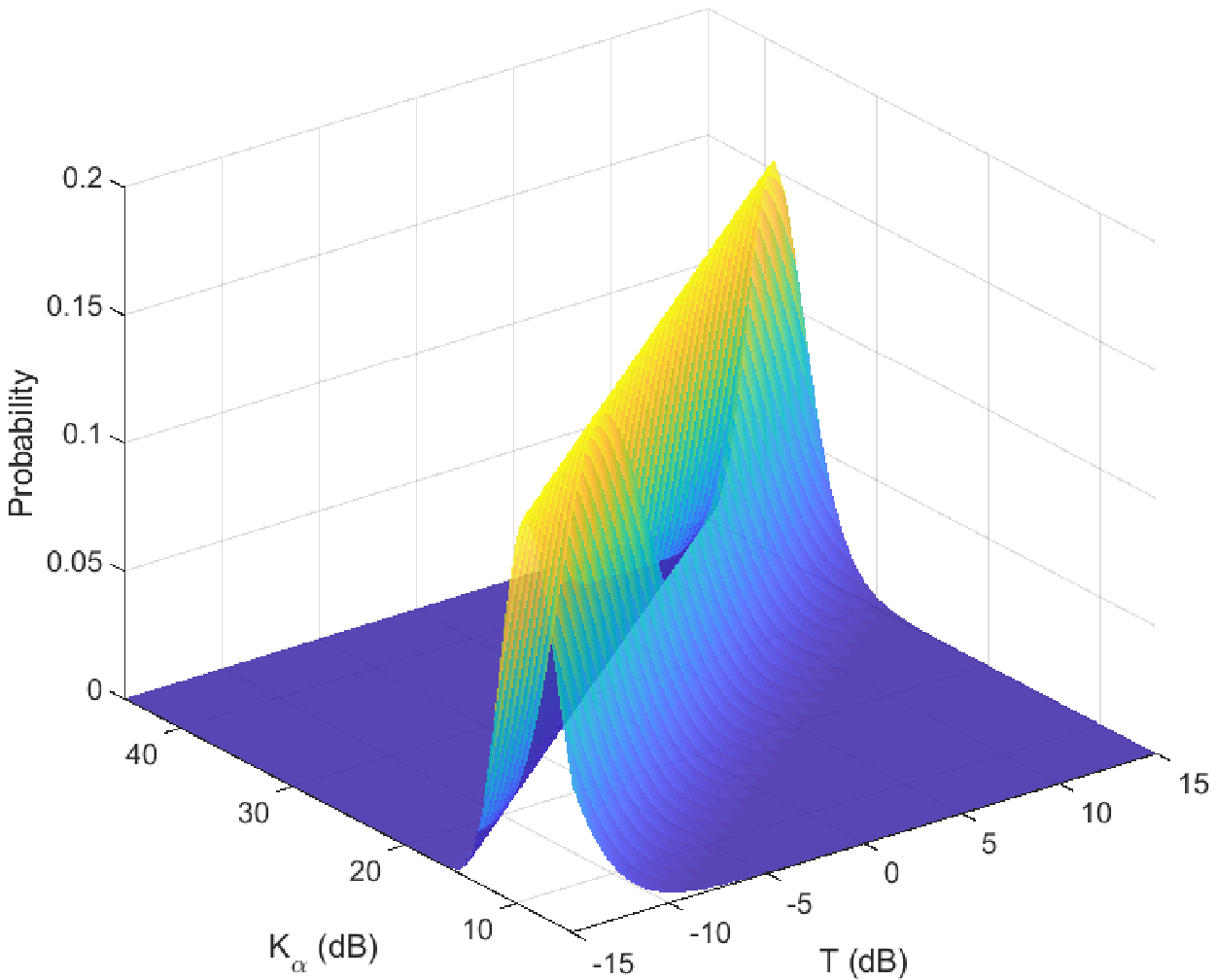}
\end{minipage}%
}%
\subfigure[\hspace{-0.0cm}]{\label{fig:subfig:2c}
\begin{minipage}[t]{2.25in}
\centering
\includegraphics[width=2.25in]{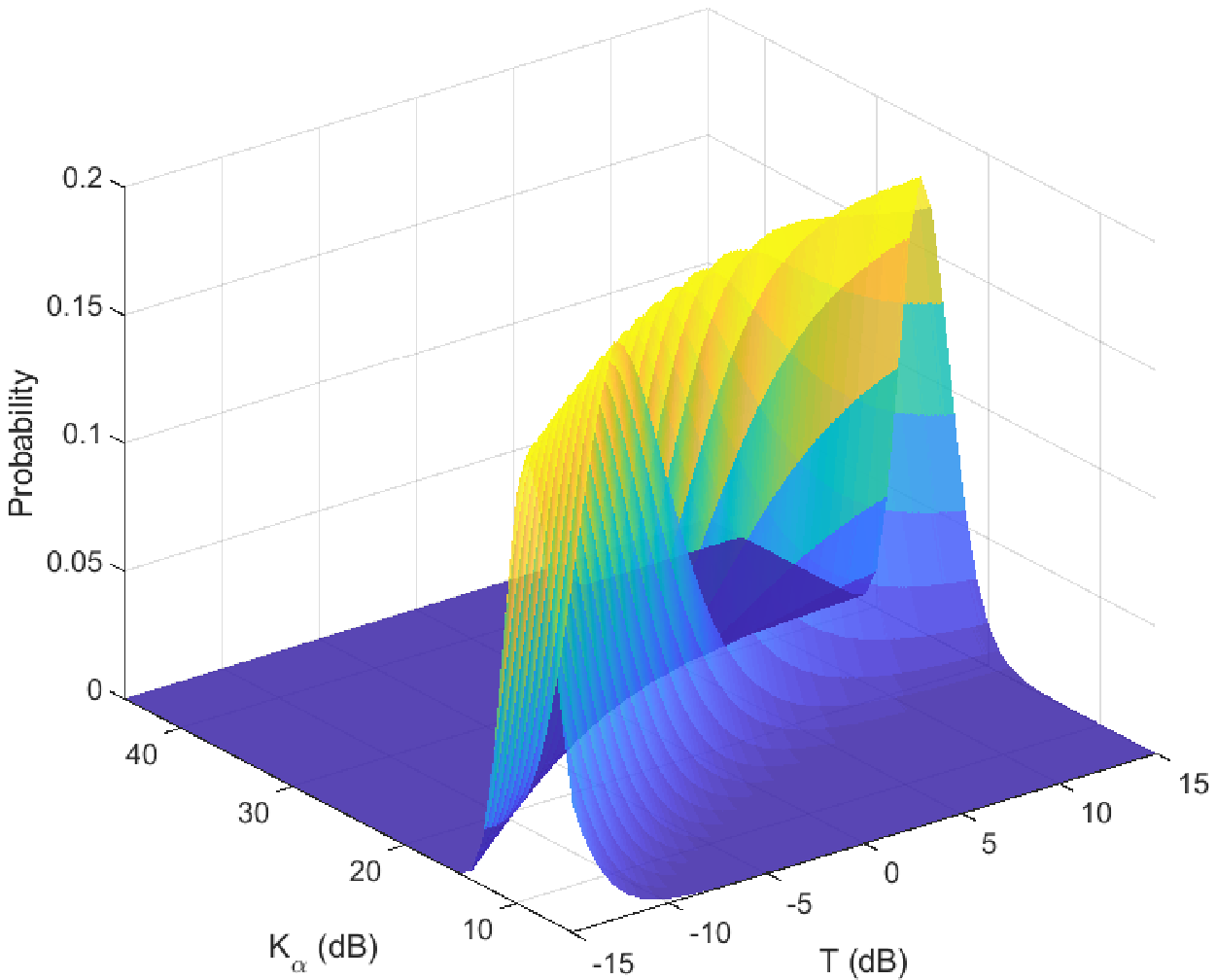}
\end{minipage}
}%
\quad
\subfigure[\hspace{-0.0cm}]{\label{fig:subfig:2d}
\begin{minipage}[t]{2.25in}
\centering
\includegraphics[width=2.25in]{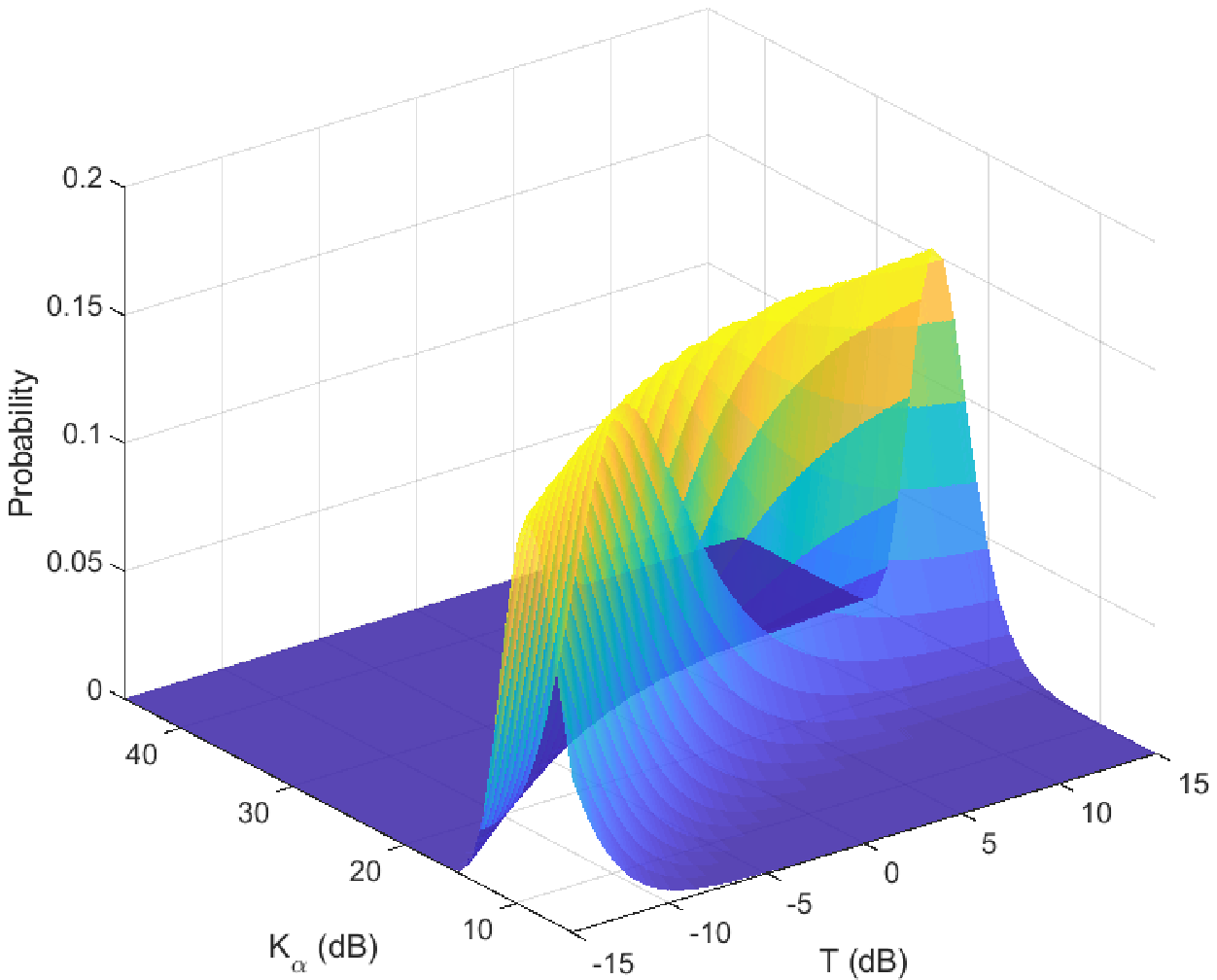}
\end{minipage}
}%
\subfigure[\hspace{-0.0cm}]{\label{fig:subfig:2e}
\begin{minipage}[t]{2.25in}
\centering
\includegraphics[width=2.25in]{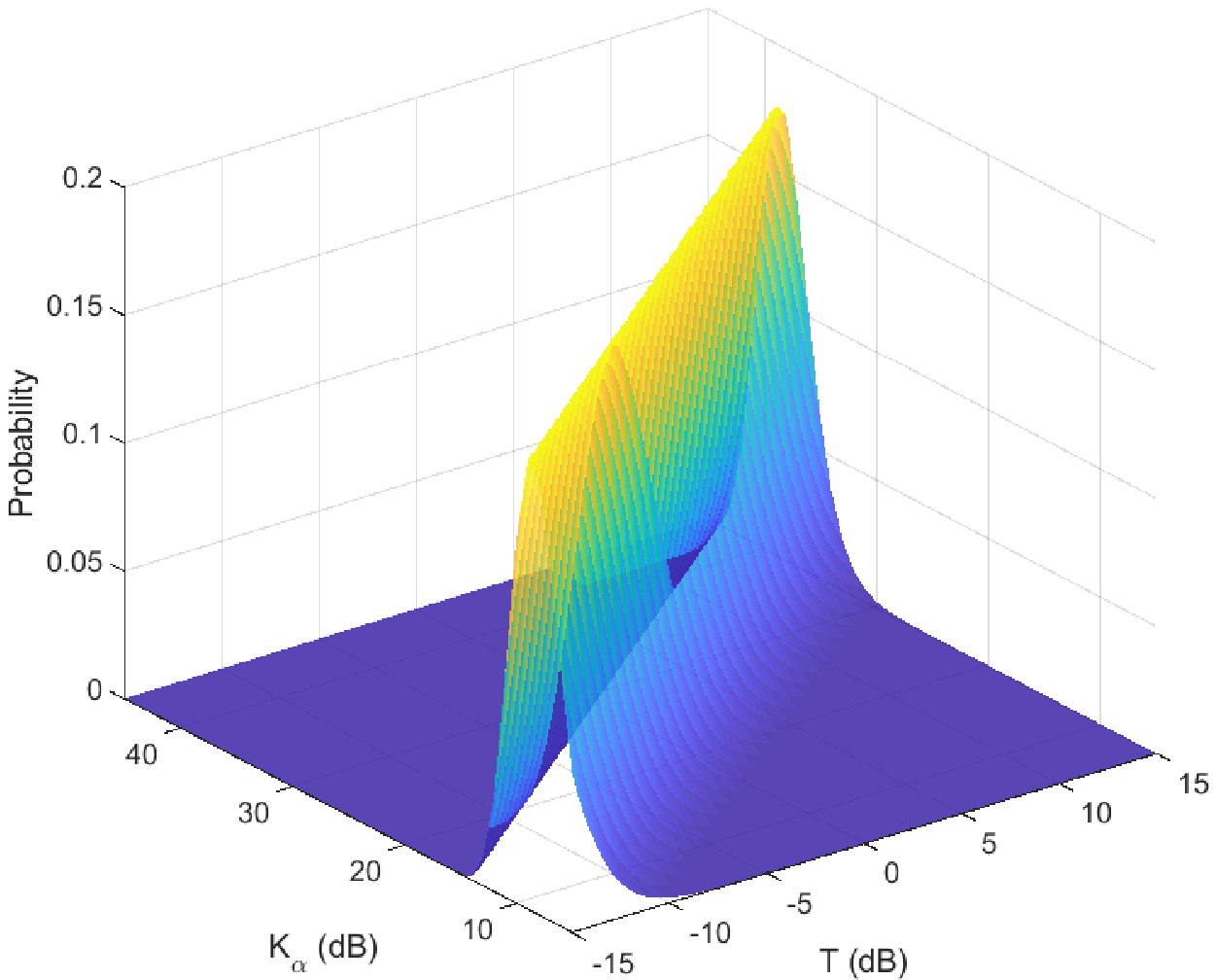}
\end{minipage}
}%
\subfigure[\hspace{-0.0cm}]{\label{fig:subfig:2f}
\begin{minipage}[t]{2.25in}
\centering
\includegraphics[width=2.25in]{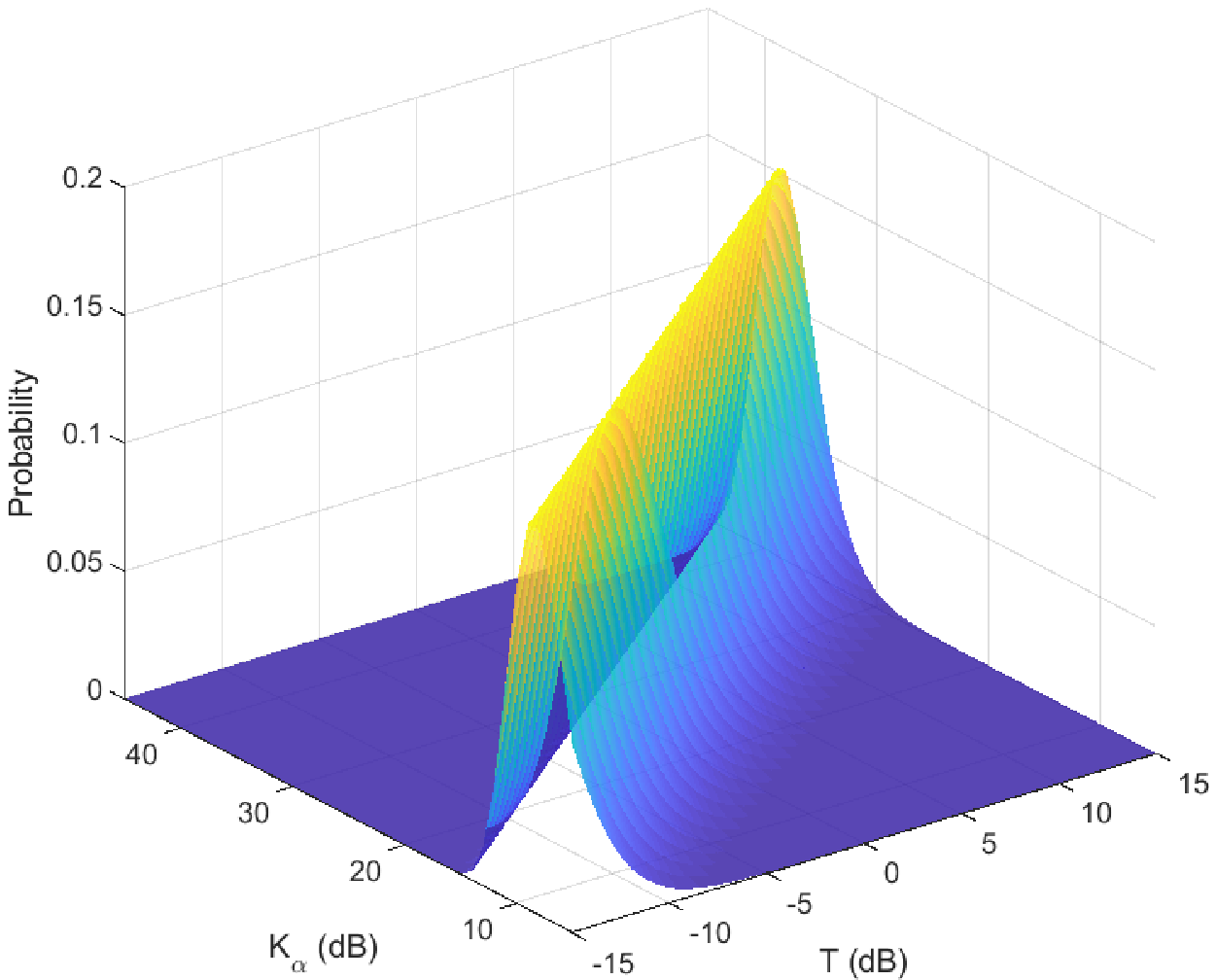}
\end{minipage}
}%
\centering
\caption{The PDF of ${K_\alpha }$ versus SNR with parameters $\left\{ {r,M,N,{{SF}},\sigma _e^2} \right\} = $(a)$\left\{ {1,2,2,7,0} \right\}$, (b)$\left\{ {0.5,3,1,7,0} \right\}$, (c)$\left\{ {1,2,2,7,0.01} \right\}$, (d)$\left\{ {0.5,3,1,7,0.01} \right\}$, (e)$\left\{ {1,2,2,7,1/\left( {1 + {L_p}{{\rm T}_{eff}}} \right)} \right\}$, (f)$\left\{ {0.5,3,1,7,1/\left( {1 + {L_p}{{\rm T}_{eff}}} \right)} \right\}$, where ${L_p}$ is set to $4$.}
\label{fig:fig2}  
\vspace{-0mm}
\end{figure*}
\begin{align}
  {P_b} &= \frac{{{2^{SF - 1}}}}{{{{\text{2}}^{SF}} - 1}}\Pr \left[ {\mathop {\max }\limits_{i,i \ne p} \left( {{Z_{{{\tilde x}_1},i}}} \right) > {Z_{{{\tilde x}_1},p}}} \right] \hfill \nonumber\\
  &{\text{   }} = \frac{{{2^{SF - 1}}}}{{{{\text{2}}^{SF}} - 1}}\int_0^\infty  {\left[ {1 - \Pr \left[ {{Z_{{{\tilde x}_1},p|X}} > \mathop {\max }\limits_{i,i \ne p} \left( {{Z_{{{\tilde x}_1},i|X}}} \right)} \right]} \right]}  \hfill \nonumber\\
  &{\text{       }} \times {f_X}\left( X \right)dX \hfill \nonumber\\
  &{\text{   }} = \frac{{{2^{SF - 1}}}}{{{{\text{2}}^{SF}} - 1}}\int_0^\infty  {\int_0^\infty  {\left[ {1 - {{\left[ {{F_{Ray}}\left( {\alpha |X;{\sigma _\beta }} \right)} \right]}^{J - 1}}} \right.} }  \hfill \nonumber\\
  &{\text{       }}\left. { \times {{\left[ {{F_{Ray}}\left( {\alpha |X;{\sigma _\tau }} \right)} \right]}^{{2^{SF}} - J}}} \right] \times {f_{Ri}}\left( {\alpha |X;{m_\alpha },{\sigma _\alpha }} \right) \hfill \nonumber\\
  &{\text{       }} \times {f_X}\left( X \right)d\alpha dX, \hfill
  \label{eq:14func}
\end{align}
where ${{f}_{X}}\left( X \right)$ is the PDF of $X$. To simplify Eq.~(\ref{eq:14func}), another equivalent form is utilized to represent ${{P}_{b}}$ \cite{8903531,8581011}, in which the error probability is expressed in terms of the noise-driven probability and the IAI-driven probability, respectively, i.e.,
\begin{align}
{{P}_{b}}=\frac{{{2}^{{SF}-1}}}{{{2}^{{SF}}}-1}\left[ P_{err}^{N}+\left( 1-P_{err}^{N} \right)\times P_{err}^{IAI} \right],
\label{eq:15func}
\end{align}
where
\begin{align}
  &P_{err}^N = \int_0^\infty  {\left[ {1 - \Pr \left[ {{Z_{{{\tilde x}_1},p|X}} > \mathop {\max }\limits_{i,i \ne p,{s_2},...,{s_J}} \left( {{Z_{{{\tilde x}_1},i|X}}} \right)} \right]} \right]}  \hfill \nonumber\\
  & \times {f_X}\left( X \right)dx \hfill \nonumber\\
  & = \int_0^\infty  {\int_0^\infty  {\left[ {1 - {{\left[ {{F_{Ray}}\left( {\alpha |X;{\sigma _J}} \right)} \right]}^{{2^{SF}} - 1}}} \right]} }  \times {f_{Ri}}\left( {\alpha ;{m_\alpha },{\sigma _\alpha }} \right) \hfill \nonumber\\
   &\times {f_X}\left( X \right)d\alpha dX, \hfill
   \label{eq:16func}
\end{align}
and
\begin{align}
  &P_{err}^{IAI} = \int_0^\infty  {\left[ {1 - \Pr \left[ {{Z_{{{\tilde x}_1},p|X}} > \mathop {\max }\limits_{i,i \ne p,i = {s_2},...,{s_J}} \left( {{Z_{{{\tilde x}_1},i|X}}} \right)} \right]} \right]}  \hfill \nonumber\\
   &\times {f_X}\left( X \right)dX \hfill \nonumber\\
   &= \int_0^\infty  {\int_0^\infty  {\left[ {1 - {{\left[ {{F_{Ray}}\left( {\alpha |X;{\sigma _\beta }} \right)} \right]}^{J - 1}}} \right]} }  \times {f_{Ri}}\left( {\alpha ;{m_\alpha },{\sigma _\alpha }} \right) \hfill \nonumber\\
   &\times {f_X}\left( X \right)dX. \hfill
   \label{eq:17func}
\end{align}
In Eqs.~(\ref{eq:16func}) and (\ref{eq:17func}), because ${{f}_{Ri}}\left( \alpha |X;{{m}_{\alpha }},{{\sigma }_{\alpha }} \right)$ in the integrand contains a modified Bessel function \cite{Watson1952A}, the overflow and accuracy problems will be faced when the numerical calculation is performed. Considering that when the Rice factor $K$ is large, the Rice distribution can be well approximated as a Gaussian distribution \cite{John1983Digital}, thus the Rice factor ${{K}_{\alpha }}$ of ${{f}_{Ri}}\left( \alpha |X;{{m}_{\alpha }},{{\sigma }_{\alpha }} \right)$ is further analyzed, which can be expressed as
\begin{align}
  {K_\alpha } = \frac{{m_\alpha ^2}}{{2\sigma _\alpha ^2}} = \frac{{X \cdot {\rm T} \cdot {2^{{{SF}}}}}}{{\left( {\sigma _e^2 \cdot {\rm T} \cdot {2^{{{SF}}}} + rM} \right)}}, \hfill
  \label{eq:18func}
\end{align}
where ${\rm T} = {E_s}/({N_0} \cdot {2^{{{SF}}}})$ is the SNR in the LoRa communication \cite{8392707}. It can be observed from Eq.~(\ref{eq:18func}) that since $X$ is a random variable, ${K_\alpha }$ is also a random variable related to $X$.
Fig.~\ref{fig:fig2} shows the PDF of ${K_\alpha }$ by Monte Carlo simulation under different parameters (e.g., $r$, $M$, $N$, ${{SF}}$, ${\rm T}$, and $\sigma _e^2$), where $\sigma _e^2 = 0$ and $\sigma _e^2 \ne 0$ correspond to the perfect CSI and the imperfect CSI scenarios, respectively, where $L_p$ denotes the number of the pilot symbol. When $\sigma _e^2$ is nonzero and fixed, it corresponds to CEEM I, while when $\sigma _e^2 = 1/\left( {1 + {L_p}{{\rm T}_{eff}}} \right)$, it corresponds to CMME II \cite{1512148,8007277}.\footnote{${{\rm T}_{eff}} = {2^{{{SF}}}} \cdot {\rm T}$ is the effective SNR for the target symbol \cite{8392707}.}
It can be seen that ${K_\alpha }$ is basically distributed in the region of ${K_\alpha } \ge 10{\rm{ dB}}$ in both perfect CSI and imperfect CSI scenarios. Hence, the Rice distributed random variable $\alpha $ can approximately follows a Gaussian distribution ${\cal N}\left( {{m_\alpha },\sigma _\alpha ^2} \right)$ \cite{John1983Digital}. However, in the imperfect CSI scenario with CEEM I, it can be observed from Figs.~\ref{fig:subfig:2c} and \ref{fig:subfig:2d} that ${K_\alpha }$ hardly increases with the increase of SNR in high SNR region.
Therefore, considering that in presence of imperfect CSI, ${P_b}$ in high SNR region is dominated by $P_{err}^{IAI}$, the Gaussian distribution will not be utilized to approximate the Rice distribution in the derivation of $P_{err}^{IAI}$ in order to ensure the accuracy of the performance analysis in high SNR region (Although this phenomenon does not occur in the CEEM II, for consistency of the derivation of $P_{err}^{IAI}$ under the CEEM I and the CEEM II, the Gaussian distribution will not be utilized to approximate the Rice distribution in the entire derivation of $P_{err}^{IAI}$).

In addition to the above numerical problems, ${{SF}} \!\in\! \left\{ {7,8,...,12} \right\}$, ${\left[ {{F_{Ray}}\left( {\alpha |X;{\sigma _J}} \right)} \right]^{{2^{{{SF}}}} - 1}}$ in Eq.~(\ref{eq:16func}) has a very high complexity.
Here, let ${\tau _{\max }} = {\max _{i,i \ne p,{s_2},...,{s_J}}}\left( {{Z_{{{\tilde x}_1},i}}} \right)$, and then refer to the procedure in \cite{8392707} to approximate the distribution of variable ${\tau _{\max }}$ by a Gaussian distribution ${\cal N}\left( {{\mu _{{\tau _{\max }}}},\sigma _{{\tau _{\max }}}^2} \right)$. Specifically, ${\mu _{{\tau _{\max }}}}$ and $\sigma _{{\tau _{\max }}}^2$ are respectively given by
\begin{align}
{\mu _{{\tau _{\max }}}} = {\left[ {{X^2}\left( {N_0^2 \cdot \hbar _{{2^{{{SF}}}} - J}^2 - {{N_0^2 \cdot {\mathchar'26\mkern-10mu\lambda _{{2^{{{SF}}}} - J}}} \over 2}} \right)} \right]^{{1 \over 4}}},
\label{eq:19func}
\end{align}
\begin{align}
  \sigma _{{\tau _{\max }}}^2 = &{X^2}\left( {{N_0}{\hbar _{{2^{{{SF}}}} - J}}}
 { - \sqrt {N_0^2\hbar _{{2^{{{SF}}}} - J}^2 - \frac{{N_0^2{\mathchar'26\mkern-10mu\lambda _{{2^{{{SF}}}} - J}}}}{2}} } \right), \hfill
  \label{eq:20func}
\end{align}
where ${\hbar _k} = \sum\nolimits_{q = 1}^k {{1 \over q}} $ denotes the ${k^{th}}$ harmonic number and ${\mathchar'26\mkern-10mu\lambda _k} = \sum\nolimits_{q = 1}^k {{1 \over {{q^2}}}} $.
\begin{figure}[tbp]
\center
\includegraphics[width=3.2in,height=2.56in]{{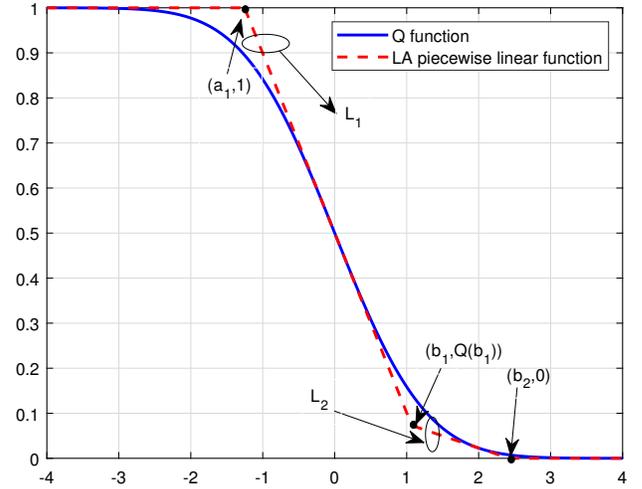}}
\caption{$Q$-function and its approximated piecewise linear function.}
\label{fig:fig3}  
\vspace{-0mm}
\end{figure}

Based on the above approximation of the distribution of the random variables $\alpha $ and ${\tau _{\max }}$, $P_{err}^N$ can be approximated as
\begin{align}
  &P_{err}^N \approx \int_0^\infty  {Q\left( {\frac{{{\mu _\alpha } - {\mu _{{\tau _{\max }}}}}}{{\sqrt {\sigma _\alpha ^2 + \sigma _{{\tau _{\max }}}^2} }}} \right)}  \cdot {f_X}\left( X \right)dX \hfill \nonumber \\
  &= \int_0^\infty  {Q\left( {\left( {\sqrt {\frac{{{X^2}{E_s}}}{{{N_0}rM}}}  - {{\left( {\hbar _{{2^{{{SF}}}} - J}^2 - \frac{{{\mathchar'26\mkern-10mu\lambda _{{2^{{{SF}}}} - J}}}}{2}} \right)}^{\frac{1}{4}}}} \right)} \right.}  \hfill \nonumber \\
  &\left. { \div \sqrt {\left( {\frac{{\sigma _e^2{E_s}}}{{2{N_0}rM}} + \frac{1}{2}} \right) + {\hbar _{{2^{{{SF}}}} - J}} - \sqrt {\hbar _{{2^{{{SF}}}} - J}^2 - \frac{{{\mathchar'26\mkern-10mu\lambda _{{2^{{{SF}}}} - J}}}}{2}} } } \right) \hfill \nonumber \\
   &\times {f_X}\left( X \right)dX, \hfill
   \label{eq:21func}
\end{align}
where $Q\left( \cdot  \right)$ is the $Q$-function \cite{John1983Digital}. Since ${SF}\in \left\{ 7,...,12 \right\}$, ${\hbar _{{2^{{{SF}}}} - J}} \gg \frac{{{\mathchar'26\mkern-10mu\lambda _{{2^{{{SF}}}} - J}}}}{2}$. Hence, one can obtain
\begin{align}
P_{err}^N \approx \int_0^\infty  Q \left( {\frac{{A\sqrt X  - B}}{C}} \right) \cdot D{X^{MN - 1}} \cdot {e^{ - Ex}}dX,
\label{eq:22func}
\end{align}
where $A = \sqrt {\frac{{{\rm T} \cdot {2^{{{SF}}}}}}{{rM}}} $, $B = \sqrt {{\hbar _{{2^{{{SF}}}} - J}}} $, $C = \sqrt {\frac{{\sigma _e^2{\rm T} \cdot {2^{{{SF}}}}}}{{2rM}} + \frac{1}{2}} $, $D = \frac{1}{{\Gamma \left( {MN} \right) \cdot {{\left( {1 + \sigma _e^2} \right)}^{MN}}}}$, $\Gamma \left(  \cdot  \right)$ denotes Gamma function \cite{Gradshteyn2007Table}, and $E =  - \frac{1}{{1 + \sigma _e^2}}$. Next, further derivations of $P_{err}^N$ and $P_{err}^{IAI}$ are performed in the perfect CSI and the imperfect CSI scenarios.
\subsection{Perfect CSI}\label{sect:perfect csi}
Since there is no IAI in the perfect CSI scenario, the BER of the proposed STBC-MIMO LoRa system only depends on the noise-driven error probability in this scenario.
Since Eq.~(\ref{eq:22func}) has no closed-form expression, here a linear approximation (LA) approach is utilized to approximate the $Q$-function as a piecewise linear function (as shown in Fig.~\ref{fig:fig3}). Accordingly, the approximated piecewise linear function of $Q\left( {\frac{{A\sqrt X  - B}}{C}} \right)$ can be expressed as
\begin{align}
Q\left( {\frac{{A\sqrt X  - B}}{C}} \right) \simeq \left\{ \begin{gathered}
  1\ \ \,\,\,X \in \left( { - \infty ,{a_1}} \right] \hfill \\
  {L_1}\ \;X \in \left( {{a_1},{b_1}} \right] \hfill \\
  {L_2}\ \;X \in \left( {{b_1},{b_2}} \right] \hfill \\
  0\ \ \,\,\,X \in \left( {{b_2}, + \infty } \right) \hfill \\
\end{gathered}  \right.,
\label{eq:23func}
\end{align}
where ${L_1}$, ${L_2}$, ${a_1}$, ${b_1}$, and ${b_2}$ are respectively expressed as
\begin{align}
{L_1} = \frac{{ - {A^2}X}}{{\sqrt {8\pi } BC}} + \left( {\frac{B}{{\sqrt {8\pi } C}} + \frac{1}{2}} \right),
\label{eq:24func}
\end{align}
\begin{align}
{L_2} = \frac{{ - {A^2}X}}{{e\sqrt {8\pi } C\left( {B + 2C} \right)}} + \frac{{\left( {B + 2C} \right)}}{{e\sqrt {8\pi } C}} + Q\left( 2 \right),
\label{eq:25func}
\end{align}
\begin{align}
{a_1} = \frac{{2{B^2} - \sqrt {8\pi } BC}}{{2{A^2}}},
\label{eq:25add1func}
\end{align}
\begin{align}
  {b_1}& = \left[ {\frac{{\left( {B + 2C} \right) - eB}}{{e\sqrt {8\pi } C}} + Q\left( 2 \right) - \frac{1}{2}} \right] \hfill \nonumber\\
  &\ \ \times \frac{{e\sqrt {8\pi } BC\left( {B + 2C} \right)}}{{{A^2}\left[ {B - e\left( {B + 2C} \right)} \right]}}, \hfill
  \label{eq:25add2func}
\end{align}
\begin{align}
{b_2} = \frac{{\left[ {e\sqrt {8\pi } C\left( {B + 2C} \right) \times Q\left( 2 \right) + {{\left( {B + 2C} \right)}^2}} \right]}}{{{A^2}}}.
\label{eq:25add3func}
\end{align}
Then, the closed-form approximated average BER expression of the proposed STBC-MIMO LoRa system in the perfect CSI scenario can be obtained as
\begin{align}
  {P_{b,pcsi}} &\approx \frac{{{2^{{\text{SF}} - 1}}}}{{{2^{{\text{SF}}}} - 1}} \cdot P_{err,pcsi}^{N,Appr} \hfill \nonumber\\
   &= \frac{{{2^{{\text{SF}} - 1}}}}{{{2^{{\text{SF}}}} - 1}} \cdot D\left( {{I_1} + {I_2} + {I_3}} \right), \hfill
   \label{eq:26func}
\end{align}
Where ${I_1}$ is derived as
\begin{align}
{I_1} = \int_0^{{a_1}} {1 \cdot {X^{MN - 1}} \cdot {e^{ - Ex}}dX} ,
\label{eq:26add1func}
\end{align}
where $e = \sum\nolimits_{k = 0}^\infty  {\frac{1}{{k!}}} $ is the base of the natural logarithm.
Utilizing \cite[Eq.(2.321)]{Gradshteyn2007Table}, ${I_1}$ can be expressed as
\begin{align}
  &{I_1} = {e^{E{a_1}}}\left[ {\sum\limits_{k = 0}^{MN - 1} {\frac{{{{\left( { - 1} \right)}^k}k!\left( {\begin{array}{*{20}{c}}
  {MN - 1} \nonumber\\
  k
\end{array}} \right)}}{{{E^{k + 1}}}}}  \times {a_1}^{\left( {MN - 1 - k} \right)}} \right] \hfill \nonumber\\
   &- \left[ {\frac{{{{\left( { - 1} \right)}^{\left( {MN - 1} \right)}}\left( {MN - 1} \right)!}}{{{E^{MN}}}}} \right], \hfill
   \label{eq:26add2func}
\end{align}
Similarly, ${I_2}$ and ${I_3}$ are expressed as Eq.~(\ref{eq:26add3func}) and Eq.~(\ref{eq:26add4func}) on the next page, respectively.
\begin{figure*}[bp]
\hrulefill
\begin{align}
  {I_2} &= \int_{{a_1}}^{{b_1}} {{L_2}}  \times {X^{MN - 1}}{e^{\left( {EX} \right)}}dX \hfill \nonumber\\
   &= \frac{{ - {A^2}}}{{\sqrt {8\pi } BC}}\left[ {{e^{EX}}\left( {\sum\limits_{k = 0}^{MN} {\frac{{{{\left( { - 1} \right)}^k}k!\left( {\begin{array}{*{20}{c}}
  {MN - 1} \\
  k
\end{array}} \right)}}{{{E^{\left( {MN + 1} \right)}}}} \times {X^{\left( {MN - k} \right)}}} } \right)\left| {\begin{array}{*{20}{c}}
  {{b_1}} \\
  {{a_1}}
\end{array}} \right.} \right] + \left( {\frac{B}{{\sqrt {8\pi } C}} + \frac{1}{2}} \right) \hfill \nonumber\\
   &\times \left[ {{e^{EX}}\left( {\sum\limits_{k = 0}^{MN - 1} {\frac{{{{\left( { - 1} \right)}^k}k!\left( {\begin{array}{*{20}{c}}
  {MN - 1} \\
  k
\end{array}} \right){X^{\left( {MN - 1 - k} \right)}}}}{{{E^{MN}}}}} } \right)\left| {\begin{array}{*{20}{c}}
  {{b_1}} \\
  {{a_1}}
\end{array}} \right.} \right]. \hfill
\label{eq:26add3func}
\end{align}
\end{figure*}
\begin{figure*}[bp]
\hrulefill
\begin{align}
  {I_3} &= \int_{{b_1}}^{{b_2}} {{L_2} \cdot {X^{MN - 1}}{e^{Ex}}dX}  \hfill\nonumber \\
   &= \frac{{ - {A^2}}}{{e\sqrt {8\pi } C\left( {B + 2C} \right)}}\left[ {{e^{EX}}\left( {\sum\limits_{k = 0}^{MN} {\frac{{{{\left( { - 1} \right)}^k}k!\left( {\begin{array}{*{20}{c}}
  {MN} \\
  k
\end{array}} \right){X^{\left( {MN - k} \right)}}}}{{{E^{\left( {MN - 1} \right)}}}}} } \right)\left| {\begin{array}{*{20}{c}}
  {{b_2}} \\
  {{b_1}}
\end{array}} \right.} \right] + \left[ {\frac{{B + 2C}}{{e\sqrt {8\pi } C}} + Q\left( 2 \right)} \right] \hfill \nonumber\\
   &\times \left[ {{e^{EX}}\left( {\sum\limits_{k = 0}^{MN - 1} {\frac{{{{\left( { - 1} \right)}^k}k!\left( {\begin{array}{*{20}{c}}
  {MN - 1} \\
  k
\end{array}} \right)}}{{{E^{MN}}}}{X^{\left( {MN - 1 - k} \right)}}} } \right)\left| {\begin{array}{*{20}{c}}
  {{b_2}} \\
  {{b_1}}
\end{array}} \right.} \right]. \hfill
\label{eq:26add4func}
\end{align}
\end{figure*}
\subsection{Imperfect CSI}\label{sect:imperfect csi}
In the imperfect CSI scenario, LA is not utilized in the derivation due to accuracy problem.\footnote{When $\sigma _e^2 \ne 0$, ${a_1}$ is less than zero in the middle and high SNR regions while the lower limit of the integral in Eq.~(\ref{eq:22func}) is zero. Therefore, utilizing the LA in the imperfect CSI scenario reduces the accuracy of the analysis for the error probability $P_{err}^N$.} To obtain the closed-form expression for Eq.~(\ref{eq:22func}) in the imperfect scenario, the Gaussian-Hermite quadrature approach is utilized  to evaluate $P_{err}^N$ \cite{handbook}, which can be expressed as
\begin{align}
\int_{ - \infty }^{ + \infty } {f\left( \xi  \right)} d\xi  = \sum\limits_{\vartheta  = 1}^\rho  {{\omega _\vartheta }f\left( {{\xi _\vartheta }} \right)} {e^{\xi _\vartheta ^2}} + {O_\rho },
\label{eq:27func}
\end{align}
where $\rho $ is the number of samples and it determines the accuracy of the approximation, ${\xi _\vartheta }$ is the ${\vartheta ^{th}}$ zero point of the Hermite polynomial, and ${O_\rho }$ is the remaining term (when $\rho $ approaches infinity, ${O_\rho }$ decreases to 0), and ${\omega _\vartheta }$ is the ${\vartheta ^{th}}$ associated weight wrriten as
\begin{align}
{\omega _\vartheta } = \frac{{{2^\rho }\rho !\sqrt \pi  }}{{{\rho ^2}\Omega _{\rho  - 1}^2\left( {{\xi _\vartheta }} \right)}}.
\label{eq:28func}
\end{align}
To solve the integral in $P_{err}^N$ , some mathematical processing needs to be performed first. Utilizing variable substitution $\xi  = \ln \left( X \right)$, $P_{err}^N$ can be rewritten as
\begin{align}
P_{err}^N = D\int_{ - \infty }^{ + \infty } {Q\left( {\frac{{A\sqrt {{e^\xi }}  - B}}{C}} \right)}  \cdot {e^{\left( {MN\xi  + E{e^\xi }} \right)}}d\xi ,
\label{eq:29func}
\end{align}
Therefore, by utilizing the Gaussian-Hermite approach given in Eq.~(\ref{eq:27func}), $P_{err}^N$ in Eq.~(\ref{eq:22func}) can be approximated as
\begin{align}
P_{err,icsi}^{N,Appr}\!\!=\!\!D\!\!\cdot\!\! \sum\limits_{\vartheta =1}^{\rho }{{{\omega }_{\vartheta }}Q\!\!\left(\!\! \frac{A\sqrt{{{e}^{\xi }}}-B}{C} \right)\!\!}\cdot \!{{e}^{\left( \xi _{\vartheta }^{2}+MN{{\xi }_{\vartheta }}+E{{e}^{{{\xi }_{\vartheta }}}} \right)\!\!}},
\label{eq:30func}
\end{align}

Next, we focus on the derivation of the IAI-driven error probability as follow
\begin{align}
  &P_{err}^{IAI} = \int_0^\infty  {\int_0^\infty  {\left[ {1 - {{\left[ {1 - {e^{\frac{{ - {\alpha ^2}}}{{2\sigma _\beta ^2}}}}} \right]}^{J - 1}}} \right]} }  \hfill \nonumber\\
   &\times \frac{\alpha }{{\sigma _\alpha ^2}}{I_0}\left( {\frac{{{m_\alpha }\alpha }}{{\sigma _\alpha ^2}}} \right){e^{\frac{{ - \left( {{\alpha ^2} + m_\alpha ^2} \right)}}{{2\sigma _\alpha ^2}}}} \cdot D{X^{\left( {MN - 1} \right)}} \hfill \nonumber\\
   &\times {e^{EX}}d\alpha dX \hfill \nonumber\\
  & = D \cdot \int_0^\infty  {\int_0^\infty  {\sum\limits_{\ell  = 1}^{J - 1} {{{\left( { - 1} \right)}^{\ell  + 1}}\left( {\begin{array}{*{20}{c}}
  {J - 1} \\
  \ell
\end{array}} \right)} } }  \cdot {e^{\frac{{ - \ell {\alpha ^2}}}{{2\sigma _\beta ^2}}}} \hfill \nonumber\\
   &\times \frac{\alpha }{{\sigma _\alpha ^2}}{I_0}\left( {\frac{{{m_\alpha }\alpha }}{{\sigma _\alpha ^2}}} \right) \cdot {e^{\frac{{ - \left( {{\alpha ^2} + m_\alpha ^2} \right)}}{{2\sigma _\alpha ^2}}}} \cdot {X^{\left( {MN - 1} \right)}} \cdot {e^{EX}}d\alpha dX, \hfill
   \label{eq:31func}
\end{align}
where ${{I}_{0}}\left( \cdot  \right)$ is the Bessel function of order zero and $\left( \begin{matrix}
   {{R}_{1}}  \\
   {{R}_{2}}  \\
\end{matrix} \right)=\frac{{{R}_{1}}!}{{{R}_{2}}!\left( {{R}_{1}}-{{R}_{2}} \right)!}$. Owing to $\sigma _{\alpha }^{2}=\sigma _{\beta }^{2}$, Eq.~(\ref{eq:31func}) becomes
\begin{align}
  P_{err}^{IAI} &= D \cdot \sum\limits_{\ell  = 1}^{J - 1} {{{\left( { - 1} \right)}^\ell }} \left( {\begin{array}{*{20}{c}}
  {J - 1} \\
  \ell
\end{array}} \right) \cdot {e^{ - \frac{{\ell m_\alpha ^2}}{{2\left( {\ell  + 1} \right)\sigma _\alpha ^2}}}} \hfill \nonumber\\
   &\times \int_0^\infty  {\int_0^\infty  {\frac{\alpha }{{\sigma _\alpha ^2}}{I_0}\left( {\frac{{{m_\alpha }\alpha }}{{\sigma _\alpha ^2}}} \right)} }  \cdot {e^{ - \frac{{\left( {\ell  + 1} \right){\alpha ^2} + \frac{{m_\alpha ^2}}{{\left( {\ell  + 1} \right)}}}}{{2\sigma _\alpha ^2}}}} \hfill \nonumber\\
   &\times {X^{\left( {MN - 1} \right)}} \cdot {e^{EX}}d\alpha dX, \hfill
   \label{eq:32func}
\end{align}
By introducing substitutions of variables ${\alpha }'=\frac{\alpha }{\sqrt{\ell +1}}$ and ${{{m}'}_{\alpha }}={{m}_{\alpha }}\sqrt{\ell +1}$, Eq.~(\ref{eq:32func}) can be rewritten as \cite{sensosloraber,John1983Digital}
\begin{align}
 & P_{err}^{IAI} = D \cdot \sum\limits_{\ell  = 1}^{J - 1} {{{\left( { - 1} \right)}^{\ell  + 1}}} \left( {\begin{array}{*{20}{c}}
  {J - 1} \\
  \ell
\end{array}} \right) \cdot {e^{ - \frac{{\ell m_\alpha ^2}}{{2\left( {\ell  + 1} \right)\sigma _\alpha ^2}}}} \hfill \nonumber\\
   &\times \frac{1}{{\ell  + 1}}\int_0^\infty  {\int_0^\infty  {\frac{{\alpha '}}{{\sigma _\alpha ^2}}{I_0}\left( {\frac{{\alpha '{{m'}_\alpha }}}{{\sigma _\alpha ^2}}} \right)} }  \cdot {e^{ - \frac{{{{\alpha '}^2} + {{m'}_\alpha }^2}}{{2\sigma _\alpha ^2}}}} \hfill \nonumber\\
   &\times {X^{\left( {MN - 1} \right)}} \cdot {e^{EX}}d\alpha' dX \hfill \nonumber\\
   &= D \cdot \sum\limits_{\ell  = 1}^{J - 1} {\frac{{{{\left( { - 1} \right)}^{\ell  + 1}}}}{{\ell  + 1}}} \left( {\begin{array}{*{20}{c}}
  {J - 1} \\
  \ell
\end{array}} \right) \hfill \nonumber\\
   &\times \int_0^\infty  {{e^{ - \left( {\frac{\ell }{{\ell  + 1}} \cdot \frac{{{\rm T} \cdot {2^{SF}}}}{{\sigma _e^2 \cdot {\rm T} \cdot {2^{SF}} + rM}} - E} \right)X}}}  \cdot {X^{\left( {MN - 1} \right)}}dX, \hfill
   \label{eq:33func}
\end{align}
Then, utilizing \cite[Eq.(3.351)]{Gradshteyn2007Table}, a closed-form expression of $P_{err}^{IAI}$ is given by
\begin{align}
  P_{err,clo.}^{IAI} = D \cdot \sum\limits_{\ell  = 1}^{J - 1} {\frac{{{{\left( { - 1} \right)}^{\ell  + 1}}}}{{\ell  + 1}}} \left( {\begin{array}{*{20}{c}}
  {J - 1}\nonumber \\
  \ell
\end{array}} \right)\left( {MN - 1} \right)! \hfill\nonumber \\
   \times {\left( {\frac{\ell }{{\ell  + 1}} \cdot \frac{{{\rm T} \cdot {2^{SF}}}}{{\sigma _e^2 \cdot {\rm T} \cdot {2^{SF}} + rM}} - E} \right)^{ - MN}}. \hfill
   \label{eq:34func}
\end{align}
Finally, combining Eqs.~(\ref{eq:15func}), (\ref{eq:30func}), and (\ref{eq:34func}), the closed-form approximated average BER expression of the proposed STBC-MIMO LoRa system in the imperfect CSI scenario can be expressed as
\begin{align}
{P_{b,icsi}}\!\!\approx \!\!\frac{{{2^{{{SF}} - 1}}}}{{{2^{{{SF}}}} - 1}}\!\!\left[ {P_{err,icsi}^{N,Appr}\!\! +\!\! \left( {1 \!\!- \!\!P_{err,icsi}^{N,Appr}} \right) \times P_{err,clo.}^{IAI}} \right].
\label{eq:35func}
\end{align}
\subsection{Analysis of diversity order}\label{sect:analysis of diversity order}
In this subsection, for gaining more insights from the BER analysis, we investigate the asymptotic BER performance in high SNR region to analyze the diversity order of the proposed STBC-MIMO LoRa system.
In the perfect CSI scenario, substituting $\sigma _e^2 = 0$ into Eq.(\ref{eq:22func}), the integral-form BER expression is given by
\begin{align}
  &{P_{b,pcsi}} = \frac{{{2^{{{SF}} - 1}}}}{{{2^{{{SF}}}} - 1}} \cdot \frac{1}{{2\Gamma \left( {MN} \right)}} \hfill \nonumber \\
  &\times \int_0^\infty  {Q\left( {\sqrt {\frac{{{\rm T} \cdot {2^{{{SF}}}}}}{{rM}} \cdot X}  - B} \right)}  \cdot {X^{\left( {MN - 1} \right)}}{e^{ - X}}dX, \hfill
  \label{eq:36func}
\end{align}
In high SNR region, ${\rm T}$ has a very large value, thus Eq.~(\ref{eq:36func}) can be approximated as
\begin{align}
  &{P_{b,pcsi,s}} \approx \frac{{{2^{{{SF}} - 1}}}}{{{2^{{{SF}}}} - 1}} \cdot \frac{1}{{2\Gamma \left( {MN} \right)}} \hfill\nonumber \\
  &\times\!\! \int_0^\infty  {Q\left( {\sqrt {\frac{{{\rm T} \!\!\cdot\!\! {2^{{{SF}}}}}}{{rM}} \cdot X} } \right)}  \cdot {X^{\left( {MN - 1} \right)}}{e^{ - X}}dX, \hfill
   \label{eq:37func}
\end{align}
where ${P_{b,pcsi}} \propto {P_{b,pcsi,s}}$. To solve the integral in Eq.~(\ref{eq:37func}), the following integral function can be employed \cite{10.5555/525693}
\begin{align}
  &\int_0^\infty  {Q\left( {\sqrt {\varpi x} } \right){x^{\varphi  - 1}}{e^{ - \frac{x}{\varsigma }}}} dx \hfill \nonumber\\
  &= \frac{1}{2}{\varsigma ^\varphi }\Gamma \left( \varphi  \right)\left[ {1 - \sum\limits_{k = 0}^{\varphi  - 1} {\mu \left( {\frac{{1 - {\mu ^2}}}{4}} \right)\left( {\begin{array}{*{20}{c}}
  {2k}\\
  k
\end{array}} \right)} } \right], \hfill
\label{eq:38func}
\end{align}
where $\mu  = \sqrt {\left( {\varpi \varsigma } \right)/\left( {2 + \varpi \varsigma } \right)} $. Hence, ${P_{b,pcsi,s}}$ in Eq.~(\ref{eq:37func}) can be written as
\begin{align}
  &{P_{b,pcsi,s}} = \frac{{{2^{{{SF}} - 1}}}}{{{2^{{{SF}}}} - 1}} \cdot \frac{1}{{\Gamma \left( {MN} \right)}} \hfill\nonumber \\
  & \times \frac{1}{2}\Gamma \left( {MN} \right)\left[ {1 - \sum\limits_{k = 0}^{MN - 1} {{\mu _1}{{\left( {\frac{{1 - \mu _1^2}}{4}} \right)}^k}\left( {\begin{array}{*{20}{c}}
  {2k} \\
  k
\end{array}} \right)} } \right], \hfill
\label{eq:39func}
\end{align}
where ${\mu _1} = \sqrt {\left( {{\rm T} \cdot {2^{{{SF}} + 1}}} \right)/\left( {2rM + {\rm T} \cdot {2^{{{SF}} + 1}}} \right)} $. According to \cite{7604059,8036271}, Eq.~(\ref{eq:39func}) can be rewritten and further simplified as
\begin{align}
  &{P_{b,pcsi,s}} = \frac{{{2^{{{SF}} - 1}}}}{{{2^{{{SF}}}} - 1}} \cdot \frac{{\Gamma \left( {MN - 1} \right)}}{{\Gamma \left( {MN} \right)}} \hfill\nonumber \\
   &\times {\left( {\frac{{1 - {\mu _1}}}{2}} \right)^{MN}}\sum\limits_{k = 0}^{MN - 1} {\left( {\begin{array}{*{20}{c}}
  {MN - 1 + k}\nonumber \\
  k
\end{array}} \right)} {\left( {\frac{{1 + {\mu _1}}}{2}} \right)^k} \hfill\nonumber \\
  & \approx \frac{{{2^{{{SF}} - 1}}}}{{{2^{{{SF}}}} - 1}} \cdot \left( {\frac{{rM}}{{{2^{SF + 2}}}}} \right) \cdot {\left( {\frac{1}{{\rm T}}} \right)^{MN}} \cdot \left( {\begin{array}{*{20}{c}}
  {2MN - 1} \\
  {MN}
\end{array}} \right). \hfill
\label{eq:40func}
\end{align}
\begin{figure}[tbp]
\center
\vspace{-0.0cm}
\subfigure[\hspace{-0.8cm}]{ \label{fig:subfig:4a}
\includegraphics[width=3.2in,height=2.56in]{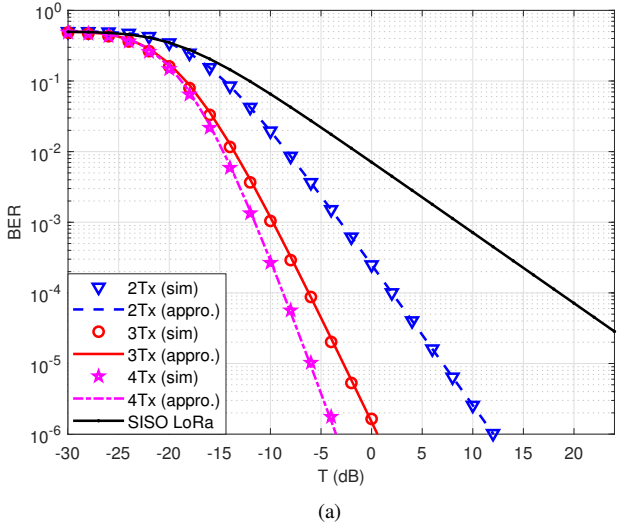}}
\vspace{-0.0cm}
\subfigure[\hspace{-0.8cm}]{ \label{fig:subfig:4b}
\vspace{-0.0cm}
\includegraphics[width=3.2in,height=2.56in]{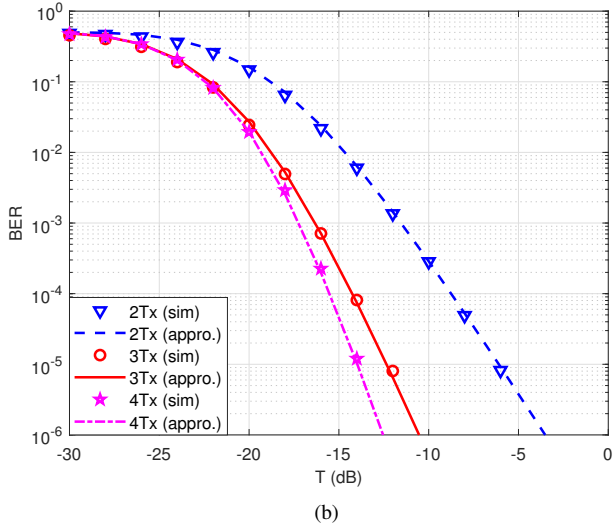}}
\caption{Simulated and approximated BER results of the proposed STBC-MIMO LoRa system with (a) 1Rx and (b) 2Rx in the perfect CSI scenario. The simulated BER performance of the SISO LoRa system is shown with black solid line in (a).}
\label{fig:fig4}
\vspace{-0mm}
\end{figure}
Hence, one can conclude from Eq.~(\ref{eq:40func}) that the diversity order of the proposed system in the perfect CSI scenario is $MN$.

Next, we study the diversity order of the proposed STBC-MIMO LoRa system in the imperfect CSI scenario. In this scenario, ${P_{b,icsi}}$ can be expressed in terms of $P_{err}^N$ and $P_{err,clo.}^{IAI}$.
For the CEEM I, when ${\rm T}$ approaches to infinity, $P_{err,icsi}^N$ and $P_{err,icsi}^{IAI}$ are approximately expressed as, respectively
\begin{align}
  &P_{err,icsi}^N \approx P_{err,icsi,s}^N = \frac{1}{{\Gamma \left( {MN} \right){{\left( {1 + \sigma _e^2} \right)}^{MN}}}} \hfill \nonumber\\
  & \times \int_0^\infty  {Q\left( {\frac{2}{{\sigma _e^2}} \cdot X} \right){X^{MN - 1}}{e^{ - \frac{X}{{1 + \sigma _e^2}}}}dX}  \hfill \nonumber\\
  & = \frac{1}{2}\left[ {1 - \sum\limits_{k = 0}^{MN - 1} {{\mu _2}\left( {\frac{{1 - \mu _2^2}}{4}} \right) \cdot \left( {\begin{array}{*{20}{c}}
  {2k} \\
  k
\end{array}} \right)} } \right], \hfill
\label{eq:41func}
\end{align}
\begin{align}
  &P_{err,icsi}^{IAI} \approx P_{err,icsi,s}^{IAI} = D \cdot \sum\limits_\ell ^{J - 1} {\frac{{{{\left( { - 1} \right)}^{\ell  + 1}}}}{{\ell  + 1}} \cdot \left( {\begin{array}{*{20}{c}}
  {J - 1} \nonumber\\
  \ell
\end{array}} \right)}  \hfill\nonumber \\
   &\times \left( {MN - 1} \right)! \times {\left( {\frac{\ell }{{\ell  + 1}} \cdot \frac{1}{{\sigma _e^2}} - E} \right)^{ - MN}} \hfill,
   \label{eq:42func}
\end{align}
where ${\mu _2} = \sqrt {\left( {1 + \sigma _e^2} \right)/\left( {1 + 2\sigma _e^2} \right)} $. Since both $P_{err,icsi,s}^N$ and $P_{err,icsi,s}^{IAI}$ are constants, the diversity order of the proposed STBC-MIMO LoRa system is zero for CEEM I. Therefore, the error floor inevitably appear in high SNR region, and one can get the expression of error floor, given by
\begin{align}
  &{P_{err\_flo.}}\!\!=\!\! \frac{{{2^{SF - 1}}}}{{{2^{SF}} - 1}}\!\! \cdot \!\!\left[ {P_{err,icsi,s}^N\! + \!\left( {1\! -\! P_{err,icsi,s}^N} \right) \times P_{err,icsi,s}^{IAI}} \right] \hfill\nonumber \\
   &\approx \frac{1}{2}\left[ {P_{err,icsi,s}^N + \left( {1 - P_{err,icsi,s}^N} \right) \times P_{err,icsi,s}^{IAI}} \right]. \hfill
\label{eq:43func}
\end{align}
According to Eq.~(\ref{eq:43func}), the error floor depends on parameters $\left\{ {M,N,\sigma _e^2} \right\}$.
For the CEEM II, as ${\rm T}$ trends to infinity, $\sigma _e^2$ becomes zero. Hence, the diversity order of the proposed system for the CEEM II is the same as that in the perfect CSI scenario, which is verified by simulations.
\begin{figure}[tbp]
\center
\includegraphics[width=3.2in,height=2.56in]{{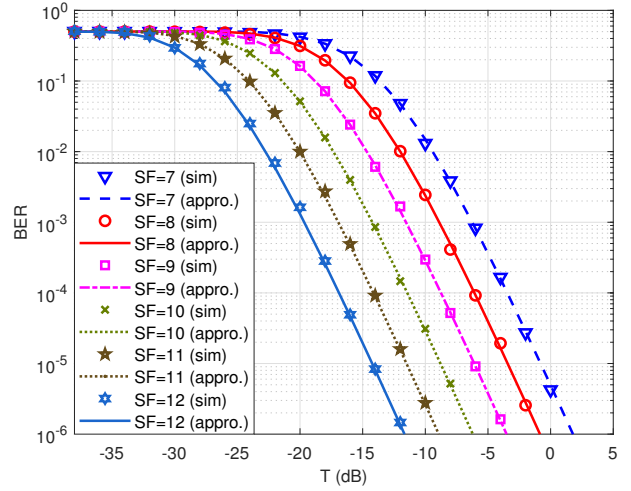}}
\vspace{-0.0cm}
\caption{Simulated and approximated BER results of the proposed STBC-MIMO LoRa system with $4{\text{Tx}}$ and $1{\text{Rx}}$ at different spreading factors in the perfect CSI scenario.}
\label{fig:fig5}  
\vspace{-0mm}
\end{figure}
\section{Simulation Results and Discussions}\label{sect:simulation results and discussion}
In this section, the performance of the proposed STBC-MIMO LoRa system with different space-time codes over a quasi-static flat Rayleigh fading channel is evaluated by simulations, and the simulation results are compared with the theoretical BER performance analysis results to verify the analysis in Sect.~\ref{sect:performance analysis}. In addition, $v{\text{Tx}}$ and $y{\text{Rx}}$ denote $v$ transmit antennas and $y$ receive antennas, respectively. In the following simulations, STBCs ${\textbf{G}_2}$, ${\textbf{G}_3}$, and ${\textbf{G}_4}$ in \cite{753730} are employed for the system with $2{\text{Tx}}$, $3{\text{Tx}}$, and $4{\text{Tx}}$, respectively.  Correspondingly, the code rates of $2{\text{Tx}}$, $3{\text{Tx}}$, and $4{\text{Tx}}$ are $1$, $0.5$, and $0.5$, respectively.
\begin{figure}[tbp]
\center
\subfigure[\hspace{-0.8cm}]{ \label{fig:subfig:6a}
\includegraphics[width=3.2in,height=2.56in]{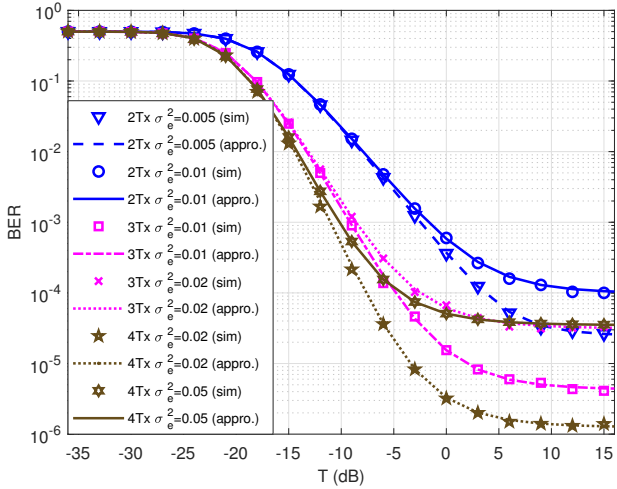}}
\subfigure[\hspace{-0.8cm}]{ \label{fig:subfig:6b}
\includegraphics[width=3.2in,height=2.56in]{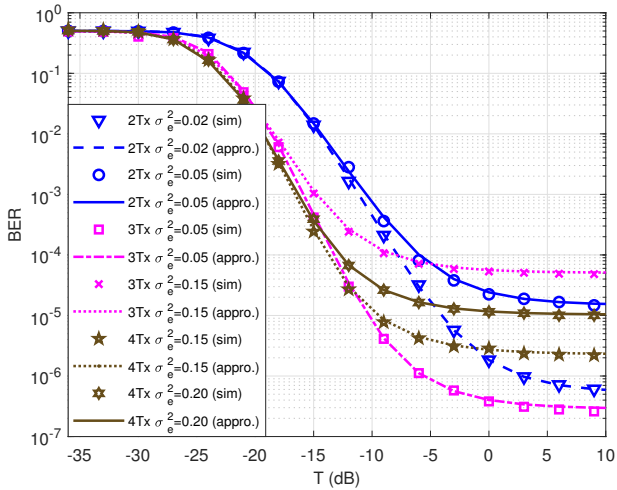}}
\vspace{0cm}
\caption{Simulated and approximated BER results of the proposed STBC-MIMO LoRa system with (a)$1{\text{Rx}}$ and (b)$2{\text{Rx}}$ in the imperfect CSI scenario (CEEM I).}
\label{fig:fig6}
\vspace{0mm}
\end{figure}

In Fig.~\ref{fig:fig4}, we plot the approximated average BER curves and simulation results in the perfect CSI scenario, where $SF$ is set to $9$. From Fig.~\ref{fig:fig4}, the derived approximated BER results can well match with the simulated ones.
Then, it can be seen from Fig.~\ref{fig:subfig:4a} that utilizing the STBC-MIMO scheme can significantly improve the diversity gain of the LoRa system. From the perspective of BER performance, taking the STBC-MIMO LoRa system with $2{\text{Tx}}$ as an instance, at a BER of ${10^{ - 4}}$, the proposed system has a 16-dB gain compared to the SISO system.
It is observed that the STBC-MIMO LoRa system with $2{\text{Rx}}$ performs better than the system with $1{\text{Rx}}$ because the former has greater diversity order than the latter.

Fig.~\ref{fig:fig5} presents the simulated and approximated BER results of the proposed STBC-MIMO LoRa system with $4{\text{Tx}}$  and $1{\text{Rx}}$ for all possible spreading factors ${{SF}} \in \left\{ {7,...,12} \right\}$ in the perfect CSI scenario. It can be seen from the figure that the simulated results are consistent with the approximated ones. In addition, BER performance increases with the increase of $SF$. This law in the STBC-MIMO LoRa system is consistent with that in the SISO LoRa system \cite{8903531}.

Furthermore, two different channel estimation error models are utilized to evaluate the proposed STBC-MIMO LoRa system with imperfect CSI, i.e., CEEM I ($\sigma _e^2$ is fixed) and CEEM II ($\sigma _e^2 = 1/\left( {1 + {L_p}{{\rm T}_{eff}}} \right)$). ${L_p}$ is set to $4$.
The simulated and approximated BER results of the proposed STBC-MIMO LoRa system in the imperfect CSI scenario with CEEM I are shown in Fig.~\ref{fig:fig6}. From this figure, it is observed that in the proposed system with the same number of the transmit and receive antenna, the error floor becomes higher as $\sigma _e^2$ increases. For a fixed $\sigma _e^2$, increasing the number of the transmit or receive antenna can reduce the error floor.

\begin{figure}[tbp]
\center
\includegraphics[width=3.2in,height=2.56in]{{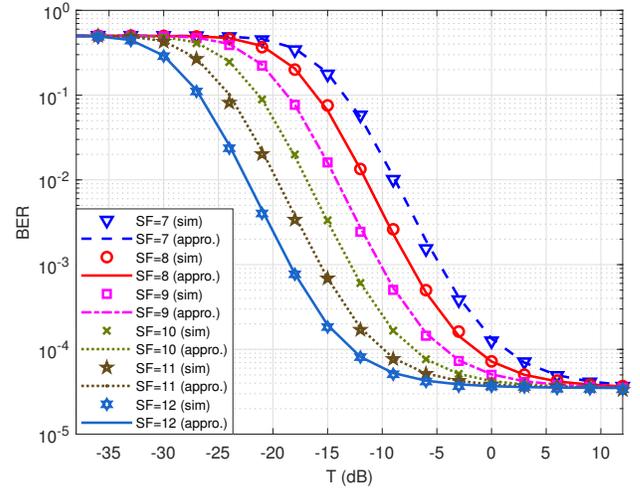}}
\caption{Simulated and approximated BER results of the proposed STBC-MIMO LoRa system with $4{\text{Tx}}$ and $1{\text{Rx}}$ in the imperfect CSI scenario (CEEM I) at different spreading factors, where $\sigma _e^2 = 0.05$.}
\label{fig:fig7}  
\vspace{-0mm}
\end{figure}
\begin{figure}[tbp]
\center
\includegraphics[width=3.35in,height=2.68in]{{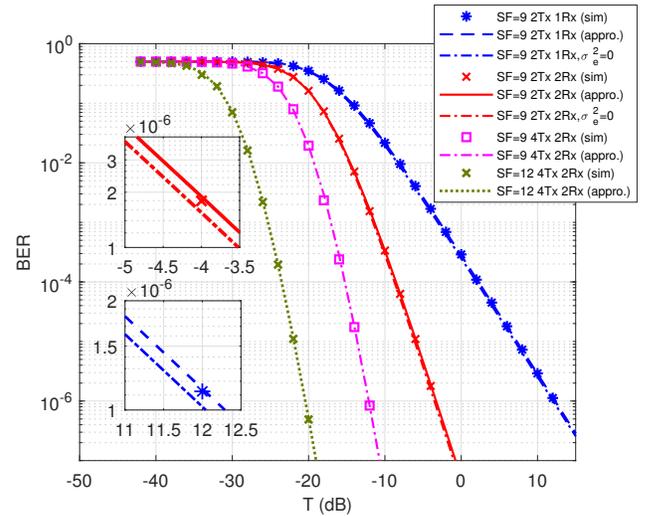}}
\caption{Simulated and approximated BER results of the proposed STBC-MIMO LoRa system in the imperfect CSI scenario (CEEM II).}
\label{fig:fig8}  
\vspace{-0mm}
\end{figure}

Fig.~\ref{fig:fig7} presents the simulated and approximated BER results of the proposed STBC-MIMO LoRa system with $4{\text{Tx}}$ and $1{\text{Rx}}$ at different spreading factors in the imperfect CSI scenario with CEEM I, where ${{SF}} \in \left\{ {7,...12} \right\}$ and $\sigma _e^2 = 0.05$. As expected, for an STBC-MIMO LoRa systems with the same parameters $\left\{ {M,N,\sigma _e^2} \right\}$, the change of ${{SF}}$ does not affect the error floor of the BER. Then, combining Fig.~\ref{fig:fig6} with Fig.~\ref{fig:fig7} not only verifies the accuracy of the derived approximated BER in the imperfect CSI scenario, but also shows that the diversity order of the proposed STBC-MIMO LoRa system is zero under the CEEM I.

Fig.~\ref{fig:fig8} shows the simulated and approximated BER results of the proposed STBC-MIMO LoRa system in the imperfect CSI scenario with CEEM II (i.e., $\sigma _e^2$ is a decreasing function of SNR). It can be seen from this figure that under the CEEM II, the diversity order of the proposed system is the same as that in the perfect CSI scenario, which verifies the conclusion obtained by the analysis of the diversity order in Sect.~\ref{sect:analysis of diversity order}. In high SNR region, BER performance of the proposed STBC-MIMO LoRa system under CEEM II is worse than that under perfect CSI, but the system performance in these two cases is very similar because $\sigma _e^2$ is very small in high SNR region. Furthermore, the results also show that the derived approximated BER expression is valid for CEEM II.
\section{Conclusions}\label{sect:Conclusions}
In this paper, an STBC-MIMO LoRa system has been presented and its theoretical performance has been carefully studied. To be specific, the closed-form approximated BER expression of the proposed STBC-MIMO LoRa system for the perfect and imperfect CSI scenarios has been derived. As a further advance, the diversity order of the proposed system has been analyzed. According to the analyzed results, the diversity order of the proposed system is zero in the imperfect CSI scenario with CEEM I, hence the error floor appears in high SNR region. In addition, full-diversity order $d=MN$ can be achieved in the imperfect CSI scenario with CEEM II and the perfect CSI scenario. Simulated results not only are in well agreement with the theoretical ones, but also verify the excellent performance and potential of the proposed STBC-MIMO LoRa system.

\end{document}